\documentclass[10pt,journal,twoside]{IEEEtran}
\usepackage{graphicx}

\usepackage{cite}
\usepackage{amsmath}
\usepackage{diagbox}
\usepackage{color}
\usepackage{verbatim}
\usepackage{subfigure}
\usepackage{float}
\usepackage{times}
\usepackage{booktabs}
\usepackage{array}
\usepackage{footnote}

\usepackage[ruled,vlined]{algorithm2e}
\usepackage{placeins}
\usepackage{tabularx,colortbl}
\usepackage{multirow}
\usepackage{makecell}
\usepackage{CJK}
\usepackage{bm}
\usepackage{amsthm}
\usepackage{extarrows}
\usepackage{amssymb}
\usepackage[table]{xcolor}
\usepackage{mathrsfs}
\usepackage{ragged2e}

\begin{document}

%

\title{
Full RGB Just Noticeable Difference (JND) Modelling}
%
%
%

\author{Jian Jin,~\IEEEmembership{Member,~IEEE,}
Dong Yu, Weisi Lin,~\IEEEmembership{Fellow,~IEEE,}
Lili Meng,
Hao Wang,
Huaxiang Zhang
\thanks{Copyright \copyright 20XX IEEE. Personal use of this material is permitted. However, permission to use this material for any other purposes must be obtained from the IEEE by sending an email to pubs-permissions@ieee.org. \emph{(Corresponding author: Weisi Lin, LiLi Meng.)}}
\thanks{J. Jin and W. Lin are with the School of Computer Science and Engineering, Nanyang Technological University, 639798, Singapore and also with Alibaba-NTU Singapore Joint Research Institute, Nanyang Technological University, 639798, Singapore. E-mail: jian.jin@ntu.edu.sg; wslin@ntu.edu.sg.}
\thanks{D. Yu, L. Meng, and H. Zhang are with the School of Information Science and Engineering, Shandong Normal University, Jinan, 250014, China. E-mail: yudong11222@163.com; mengll\_83@hotmail.com; huaxzhang@hotmail.com.}
\thanks{H. Wang is with the Alibaba cloud business group, Alibaba, Hangzhou 310052, China. Email: tianye.wh@alibaba-inc.com.}}

\markboth{Submit to IEEE Transactions on Image Processing,~Vol.~X, No.~X, Nov.~2021}%
{Jian \MakeLowercase{\textit{et al.}}:
~Full RGB Just Noticeable Difference (JND) Modelling}
%

\maketitle


\begin{abstract}
Just Noticeable Difference (JND) has many applications in multimedia signal processing, especially for visual data processing up to date. It’s generally defined as the minimum visual content changes that the human can perspective, which has been studied for decades. However, most of the existing methods only focus on the luminance component of JND modelling and simply regard chrominance components as scaled versions of luminance. In this paper, we propose a JND model to generate the JND by taking the characteristics of full RGB channels into account, termed as the RGB-JND. To this end, an RGB-JND-NET is proposed, where the visual content in full RGB channels is used to extract features for JND generation. To supervise the JND generation, an adaptive image quality assessment combination (AIC) is developed. Besides, the RDB-JND-NET also takes the visual attention into account by automatically mining the underlying relationship between visual attention and the JND, which is further used to constrain the JND spatial distribution. To the best of our knowledge, this is the first work on careful investigation of JND modelling for full-color space. Experimental results demonstrate that the RGB-JND-NET model outperforms the relevant state-of-the-art JND models. Besides, the JND of the red and blue channels are larger than that of the green one according to the experimental results of the proposed model, which demonstrates that more changes can be tolerated in the red and blue channels, in line with the well-known fact that the human visual system is more sensitive to the green channel in comparison with the red and blue ones.
\end{abstract}


%

\section{Introduction}
\subsection{Motivation}

JND reflects the perceptual redundancy in visual signals for the Human Visual System (HVS) \cite{hall1977nonlinear} due to the unique psychological and physiological mechanisms of the HVS, and is widely used in many perceptual image/video processing applications \cite{lin2021progress}, such as perceptual image/video coding \cite{wu2013perceptual,kim2015hevc,zhou2020just}, watermarking and information hiding \cite{cheng2001additive,li2019orientation}, image/video quality assessment \cite{wang2018analysis,wang2018user}, perceptual image/video enhancement \cite{cheng2018performance}, and so on. 
The existing JND models can be divided into two categories according to the way of JND modelling, namely HVS-inspired models  \cite{chou1995perceptually,yang2005motion,liu2010just,wu2013just,wu2017enhanced,qi2016stereoscopic,jia2006estimating,bae2013novel,bae2014novel,niu2013visual,hadizadeh2017saliency,zeng2019visual} and learning-based models \cite{liu2019deep,zhang2021deep,tian2021perceptual,wu2020unsupervised,jin2021just,jin2016statistical,wang2016mcl,wang2017videoset,liu2018jnd}.


The former ones model the JND by utilizing the characteristics of the HVS. For instance, background luminance \cite{bae2013novel}, contrast \cite{bae2014novel}, pattern complexity \cite{wu2017enhanced}, and visual attention (visual saliency) \cite{niu2013visual,hadizadeh2017saliency,zeng2019visual} are considered in designing various JND models. However, some of considerations are very rough, e.g., the JND is scaled by a set of handcrafted weights in \cite{niu2013visual}. According to the related investigations on the visual attention's impacts on the JND \cite{hadizadeh2017saliency}, the relation between visual attention and the JND is quite complicated, which should be represented with a more accurate model. Besides, all these the characteristics (e.g., the background luminance, contrast, pattern complexity, etc.) are used for luminance component of JND modelling and chrominance components are simply regarded as the scaled versions of luminance. Hence, the JND models above cannot accurately represent the visual redundancy of the HVS in full-color space. More explorations are needed for the HVS-inspired models, especially for impacts on the stimuli from full-color channels on JND modelling. 

With the great successes achieved by deep learning, especially in visual signal processing and understanding, some works have explored the JND with learning-based methods along two directions, namely data-driven JND modelling \cite{liu2019deep,zhang2021deep,tian2021perceptual} via supervised learning and generative JND modelling \cite{wu2020unsupervised,jin2021just} via unsupervised or weakly supervised learning. However, data-driven methods largely rely on the labeled dataset. As most of the JND datasets \cite{jin2016statistical,wang2016mcl,wang2017videoset,liu2018jnd} are built for image/video coding purposes, only limited visual data with few codecs are labeled by subjects for JND modelling so far. Such a kind of datasets cannot be used to build the JND models for the distortions caused by the other factors, e.g., transmission error, pre-/post-processing, Gaussian noise/blur, and so on. To the best of our knowledge, there are at least 17 distortions of image/video \cite{liu2012image}. Besides, each image/video may be degenerated by several kinds of distortions together. Furthermore, as new technologies/applications develop on images and videos, new kinds of distortions will be involved continuously, such as generative network caused distortions \cite{goodfellow2014generative}, style transfer caused differences \cite{gatys2016image}, and so on. Therefore, it’s impractical to build very large JND datasets to cover all kinds of distortions with different distorted levels, since data annotation is time-consuming and expensive. In view of this, the generative JND model proposed in \cite{wu2020unsupervised} generates the JND as a kind of noise, but it will not lead to the decline of Image Quality Assessment (IQA) \cite{bosse2017deep}. In other words, the generated noise will not be perceived by the HVS and will not lead to the visual quality decline to the HVS. Meanwhile, an auto-encoder based generated model \cite{jin2021just} is proposed to explore the JND of deep machine vision. However, such kind of model still takes the extracted deep features of a single channel into account instead of those of full-color channels, which cannot well represent the redundancy of the HVS in the full-color space. 

So far, there is no work on carefully modelling the JND for full-color channels. It will help us learn more about the visual redundancy of the HVS in the full-color space and provide technical support for related applications.

\subsection{Contributions}
In this paper, we attempt an initial exploration on accurately JND modelling for full-color space by considering the stimuli of full-color channels via a learning-based generative network, i.e., the proposed RGB-JND-NET with an auto-encoder like design. To this end, the handcrafted features (visual attention and pattern complexity maps) together with their associated visual content of full RGB channels (original image) are fed into the RGB-JND-NET to provide relevant information for full RGB JND (termed as \textbf{RGB-JND}) generation. Besides, an adaptive IQA combination (AIC) method is proposed to recognize and evaluate the distortion caused by the generative network to supervise the RGB-JND generation. The main contributions of this work can be summarized as follows:

\begin{itemize}
    \item This is the first work on careful investigation of JND modelling for full-color space, exploring the impacts of stimuli of full-color channels. It may catalyze positive chain-effects in color display and capturing optimization, visual signal representation/compression, visual understanding, and their relevant applications due to its fundamental nature. 
    \item To supervise the RGB-JND generation, the AIC method is proposed, which can effectively recognize and evaluate the distortion caused by the generative network. As there is no dedicated IQA metric on measuring generative network caused distortion, the AIC firstly use a well-learnt classifier to classify such distortion into the most related known distortions. Then, several IQAs, which can well measure the most related known distortion, will be adaptively combined for evaluation.
    \item Visual attention is also taken into account during our modelling. Unlike being used as the handcrafted weights of the JND in traditional methods \cite{niu2013visual,hadizadeh2017saliency,zeng2019visual}, visual attention is used as a feature and fed into the RGB-JND-NET, where the underlying relationship between the visual attention and the RGB-JND is automatically mined and utilized to constrain the RGB-JND during the RGB-JND-NET optimization. By seriatim applying the techniques above, the proposed model can accurately model the redundancy of the HVS for full RGB channels.
\end{itemize}

The outline of the rest of our paper is as follows. Section II
reviews the JND models and IQA methods. Section III presents the full RGB JND model. Section IV presents experimental results and
Section V concludes this paper.

\section{Related Works}
\label{RW}
Unlike a high-level analysis of the JND models in Section I-A, more detailed reviews on the JND models are elaborated in this section along two directions, i.e., HVS-inspired and learning-based models. Since the proposed AIC method is highly related to the IQAs, the IQAs and their relevant technologies are also reviewed in this section to make the proposed AIC easily understood by readers.

\subsection{JND review}

\subsubsection{HVS-inspired JND models} as aforementioned, such kind of models are built based on the characteristics of HVS. As previous researches, Chou et al. \cite{chou1995perceptually} first proposed a spatial-domain JND model by combining Luminance Adaptation (LA) and Contrast Masking (CM). Both the LA and CM are calculated on the luminance channel. To characterize the possible overlaps between LA and CM, Yang et al. \cite{yang2005motion} further utilized a nonlinear additivity model for masking effects (NAMM) to better represent a generalized spatial JND model. Afterwards, Liu et al. \cite{liu2010just} first proposed the Edge Masking (EM) and Texture Masking (TM), which improves the CM estimation process and further enhanced the JND model. Then, Wu et al. \cite{wu2013just} first introduced the free-energy principle in the JND modeling to achieve comprehensively evaluation. After that, they observed that there was a relationship between the Pattern Complexity (PC) \cite{wu2017enhanced} of visual content and JND, which was further used to decide the masking effects and achieve a good performance. Meanwhile, several works \cite{niu2013visual,hadizadeh2017saliency,zeng2019visual,qi2016stereoscopic} were proposed by taking the Visual Attention (VA) into account, where the VA was used as the handcrafted weights and roughly incorporated into CM during JND modelling. All these methods above can directly obtain the JND of each pixel. However, they cannot be directly used in image/video perceptual coding, as the compression is based on a transformed block, such as DCT block. In view of this, lots of transform domain JND models \cite{jia2006estimating,bae2013novel,bae2014novel} were developed. The well-known contrast sensitivity function (CSF) which reflects the characteristics of HVS in the spatial frequency domain was typically modelled. For instance, Jia et al. \cite{jia2006estimating} formulated the CSF based on the sub-band JND threshold, which was obtained by the subjects, as an exponential function of the spatial contrast during JND modelling. Considering the frequency characteristics, Bae et al. \cite{bae2013novel} proposed a new JND model based on DCT sub-band. After that, they further incorporated the CFS, LA, and CM and proposed a new sub-band JND profile \cite{bae2014novel}. However, the modelling of the JND threshold for each pixel or sub-band above is achieved separately in the pixel or sub-band domain by being summed up in a local neighborhood. Hence, it cannot reflect the total masking of the whole frame. Besides, the HVS-inspired methods are also yielded to our limited knowledge on the HVS.

\subsubsection{Learning-based JND models} in some recent works \cite{liu2019deep,zhang2021deep,tian2021perceptual,wu2020unsupervised,jin2021just,jin2016statistical,wang2016mcl,wang2017videoset,liu2018jnd}, learning based JND models were proposed via data-driven supervised learning methods \cite{liu2019deep,zhang2021deep,tian2021perceptual} and generative unsupervised learning methods \cite{wu2020unsupervised, jin2021just}, which were able to generate the JND by modelling the masking at the whole picture and even for video level. For instance, Jin et al. \cite{jin2016statistical} built the first picture-wise JND dataset, where the distortions caused by the JPEG codec with different QFs was evaluated and labeled by the subjects to find the JND point. Further, Wang et al. \cite{wang2017videoset} proposed a subjective methodology, i.e., Satisfied-User-Ratio (SUR), to find the video-wise JND videos and built the first video-wise JND dataset, namely VideoSet. After that, Liu et al. \cite{liu2019deep} made the picture-wise JND prediction as a classification problem, where deep learning technique was utilized to predict JND point for compressed images based on the MCL-JCI dataset. Then, Zhang et al. \cite{zhang2021deep} built a video-wise JND and SUR model based on deep learning techniques. It can predict the quality of compressed videos based on the VideoSet. Such kinds of models can achieve picture-wise and video-wise JND prediction via the data-driven supervising learning methods. However, they highly depend on the datasets. Manual annotation is time-consuming and expensive works. Such kind of methods cannot cover all the distortions during JND modelling. To handle this, Wu et al. \cite{wu2020unsupervised} proposed utilizing the generative networks to generate the JND via an unsurprising/weakly supervising method. They used the IQA metric to supervise the JND generation instead of subjects observing to simulate the process that the generated JND can be tolerated by the HVS. Besides, they also absorbed the PC in \cite{wu2017enhanced} as the feature to make a further improvement and achieve the state-of-the-art. Meanwhile, Jin et al. \cite{jin2021just} used the auto-encoder to generate the first JND for deep machine vision. Instead of using IQA as the supervision, they used the performance metric of the deep machine vision task to guild the JND generation, Meanwhile, they also took the attention of the neural network into account and achieved JND modelling for deep machine vision. 

As we know, there are about 10 million rod cells and 5 million cone cells in the retina, which are responsible for perceiving the luminance intensity and distinguishing the color \cite{curcio1990human}, respectively. According to such facts, lots of color spaces, such as RGB, YUV, etc., are developed and widely used in displays, cameras, and their relevant applications. However, most of the existing JND models mainly took the characteristics of single channel into account. For instance, the characteristics of luminance channel is mainly considered during JND modelling, since luminance channel is more sensitive to the HVS compared with the others. Hence, such kind of JND models ignored the impact of stimuli among full-color channels, which makes them can hardly achieve high consistency to the HVS.

\subsection{IQA review}

Image quality assessment (IQA) plays a critical role in image processing, which has been developed for decades. Nowadays, a number of IQA metrics have been proposed as better alternatives, such as MSE, PSNR, MS-SSIM \cite{wang2003multiscale}, SSIM \cite{wang2004image}, VIF \cite{sheikh2006image}, VSNR \cite{chandler2007vsnr}, NQM \cite{damera2000image}, PSNR-HVS \cite{egiazarian2006new}, IFC \cite{sheikh2005information}, FSIM \cite{zhang2011fsim}, and so on. They are developed with the same goal, i.e., reflecting the image quality as accurately as the HVS does. However, hardly IQA metrics above can achieve this, as they can only achieve good performance for certain kinds of image distortions. As shown in TABLE \ref{TableI}, there are at lest 17 distortion types. No single IQA metric that can maintain the best performance in all situations. For example, MSE and PSNR, which are widely used IQA metrics, can be used for evaluate the signal fidelity distortion. However, they still cannot keep high consistence with human perception in some cases. To overcome this, Liu et al. \cite{liu2012image} proposed multi-method fusion (MMF) proposed, where 17 distortion types are divided into five groups as shown in TABLE \ref{TableII}. After that, a combination of the multiple existing metrics via machine learning achieves a state-of-the-art accuracy. As the multimedia technologies developments, more and more distortion types are generated, such as deep learning based generative network caused distortion, style transfer caused distortion, and so on. However, there are hardly IQA metrics developed for measuring such kinds of distortion. Therefore, IQA is still an open question. 






\begin{table}[t]
\begin{center}
\tiny

    \caption{\centering{\scshape Distortion Types in TID 2008 \cite{ponomarenko2009tid2008} Dataset}}
    \setlength{\tabcolsep}{6mm}{

    \begin{tabular}{c|c}
\hline 
Type Index& Distortion Type\\
\hline  
1&Additive Gaussian noise\\
\hline  
2&Different additive noise in color components\\
\hline  
3&Spatially correlated noise\\
\hline  
4&Masked noise\\
\hline  
5&High frequency noise\\
\hline  
6&Impulse noise\\
\hline  
7&Quantization noise\\
\hline
8&Gaussian blur\\
\hline  
9&Image denoising\\
\hline  
10&JPEG compression\\
\hline  
11&JPEG2000 compression\\
\hline  
12&JPEG transmission errors\\
\hline  
13&JPEG2000 transmission errors\\
\hline  
14&Non eccentricity pattern noise\\
\hline  
15&Local block-wise distortions of different intensity\\
\hline  
16&Mean shift (intensity shift)\\
\hline  
17&Contrast change\\
\hline 
\end{tabular}}
\label{TableI}
\end{center}
\vspace{-0.6cm}
\end{table}

\begin{table}
\tiny
    \centering
        \caption{\centering{\scshape Distortion Types Are Divided into Distortion Groups}}
        \setlength{\tabcolsep}{4.4mm}{
    \begin{tabular}{c|c|c|c|c|c}
\hline 
Type Index & 1-7& 8-9 & 10-11 & 12-13 & 14-17\\
\hline  
Group Index & $1$ & $2$ & $3$ & $4$ & $5$\\
\hline  

\end{tabular}}
\label{TableII}
\vspace{-0.6cm}
\end{table}


\section{Full RGB JND Model}
\label{model}
In this section, the RGB-JND is firstly defined and formulated. After that, three main challenging issues on full RGB JND modelling are elaborated and well-solved with the proposed techniques. Finally, an RGB-JND-NET is proposed to generate the RGB-JND.

\subsection{Definition and formulation}
A good JND model is required to accurately reflect the redundancy of the HVS in different visual content. Generally, the perception of the HVS is determined by the information captured from rod and cone cells in the retina. As reviewed in Section \ref{RW}, cone cells can distinguish three primary colors (red, green, and blue). Although they have different sensitivities to different color, the perception of the HVS is an integrated result of the stimuli from three primary colors. Therefore, the JND model needs to be built for full-color space by considering the characteristics of full-color channels. \textbf{In this paper, the RGB-JND is first defined: for each pixel, there are three thresholds corresponding to RGB three channels. Any changes in RGB three channels under such thresholds will not be perceived by the HVS. Such thresholds are the RGB-JND.}

Generally, without being perceived by the HVS, the larger JND is tolerated, the better JND model is. Therefore, the RGB-JND can be formulated as an optimization problem, which is as follows:
\begin{equation}
\label{formulation1}
\begin{split}
\arg \min _{\vec{j}}\left(\alpha \cdot \frac{1}{J}+\beta \cdot \mathcal{M}(\vec{o},(\vec{o} \ \textcircled{+} \  \vec{j})) \right)\\
= \arg \min _{\vec{j}}\left(\alpha \cdot \frac{1}{J}+\beta \cdot \mathcal{M}(\vec{o},\vec{d}) \right),
\end{split}
\end{equation}
where $J$ is the magnitude of the RGB-JND. We have $J = \sum\limits^{c} \sum\limits^{h} \sum\limits^{w}|\vec{j}|$. $\vec{j}$ is the RGB-JND, which is a $c \times h \times w$ tensor, denoted by $\vec{j} \in \mathbb{R}^{c \times h \times w}$. $|\cdot|$ is the absolute operation. The first dimension $c$ ($c=3$) represents the color channel, i.e., red, green, and blue channel. $h$ and $w$ are the height and width of the RGB-JND, respectively. $\vec{j} = \{\vec{j}_r, \vec{j}_g, \vec{j}_b \}$, where $\vec{j}_r, \vec{j}_g, \vec{j}_b \in \mathbb{R}^{1 \times h \times w}$. Similarly, the original image is denoted by $\vec{o}$, where $\vec{o} \in \mathbb{R}^{c \times h \times w}$. Also, we have $\vec{o} = \{\vec{o}_r, \vec{o}_g, \vec{o}_b \}$, where $\vec{o}_r, \vec{o}_g, \vec{o}_b \in \mathbb{R}^{1 \times h \times w}$. $\vec{d}$ is the RGB-JND distorted image, and we have $\vec{d} = \vec{o} \ \textcircled{+} \  \vec{j}$, where $\textcircled{+}$ is the element-wise adding operation. Therefore, we have \begin{equation}
\label{distorted}
\vec{d} = \{ \vec{d}_r, \vec{d}_g, \vec{d}_b \} = \{\vec{o}_r \ \textcircled{+} \  \vec{j}_r, \vec{o}_g \ \textcircled{+} \  \vec{j}_g, \vec{o}_b \ \textcircled{+} \  \vec{j}_b \}. 
\end{equation}
$\mathcal{M}(\cdot, \cdot)$ is an IQA metric (e.g., the MSE). $\alpha$ and $\beta$ are two parameters to balance their associated two items in formula \eqref{formulation1}. 

In formula \eqref{formulation1}, the first item is used to constrain the magnitude of the generated RGB-JND as large as possible. Meanwhile, the second item is used to constrain that the generated RGB-JND will not be perceived by the HVS (i.e., will not lead to perceptual quality decrease of the HVS), which is a similar strategy as used in \cite{wu2020unsupervised,jin2021just}. Then, the RGB-JND become an optimization problem, which can be roughly achieved with the generative neural networks. Namely, generative neural network will automatically generate the RGB-JND with the constraints above. 

However, according to our initial attempt based on the RGB-JND formulation above, there are still a gap between the RGB-JND and the perceptual redundancy of the HVS. Therefore, the RGB-JND formulation in formula \eqref{formulation1} need to be refined. Specially, three challenging problems that need to be elaborately addressed, which are summarized as follows:
\begin{itemize}
\item Quantificationally modelling the RGB-JND magnitude of full RGB channels. For instance, the first item in formula \eqref{formulation1} only makes sure that the magnitude of the generated RGB-JND is as large as possible. A more detailed constraint of the RGB-JND magnitude need to be designed according to the visual content along three color channels. 
\item Effectively evaluating the reasonability of the generated RGB-JND. The RGB-JND is generated via the generative neural network. However, there is no IQA developed for measuring the distortion caused by generative neural network. Therefore, it hard to achieve effective evaluation of the RGB-JND using the existed IQAs. 
\item Reasonably allocating the RGB-JND for each pixel. So far, no constraints or guidance are used for RGB-JND spatial distribution during RGB-JND generation, which plays an critical role in full RGB JND modelling. 

\end{itemize}



\subsection{Gradient based magnitude constraint of full RGB channels}
As aforementioned, a good JND model should be highly consistent with the HVS. Large literature investigations show a fact that the magnitude of the visual perceptual redundancy (JND) is proportional to the complexity of the visual content. For instance, the HVS cannot perceive large noises on an image with complex texture. Whereas, little noises are noticeable, when the image has a simple texture. Generally, the gradient of the image is used to represent the texture complexity of the image, which are widely used in the image and video processing \cite{liu2010just,wu2020unsupervised}. The gradient of the image is denoted by $G$, and it can be represented as
\begin{equation}
\label{formulation2}
G = \sum\limits^{c} \sum\limits^{h} \sum\limits^{w} \sqrt{g_{1}(c,h,w)^2 + g_{2}(c,h,w)^2},
\end{equation}
where $g_{1}(c,h,w)$ and $g_{2}(c,h,w)$ are the vertical and horizontal gradient in channel $c$ of pixel $(h,w)$.  

\begin{figure}[htbp]
    \centering
    \renewcommand\thesubfigure{(a\arabic{subfigure})}
    \setcounter{subfigure}{0}
    \subfigure[]{\includegraphics[width=2.5cm]{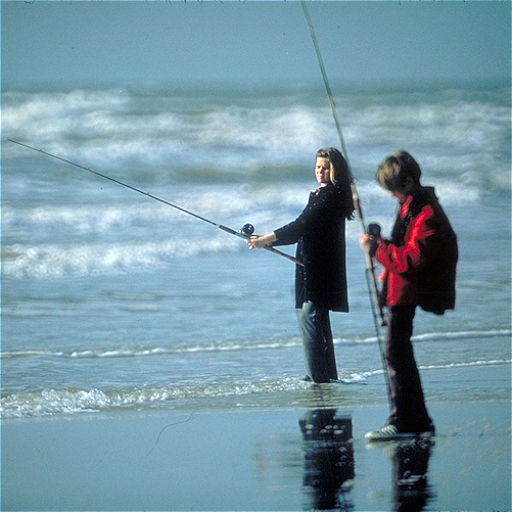}}
    \renewcommand\thesubfigure{(a\arabic{subfigure})}
    \setcounter{subfigure}{1}
    \subfigure[]{\includegraphics[width=2.5cm]{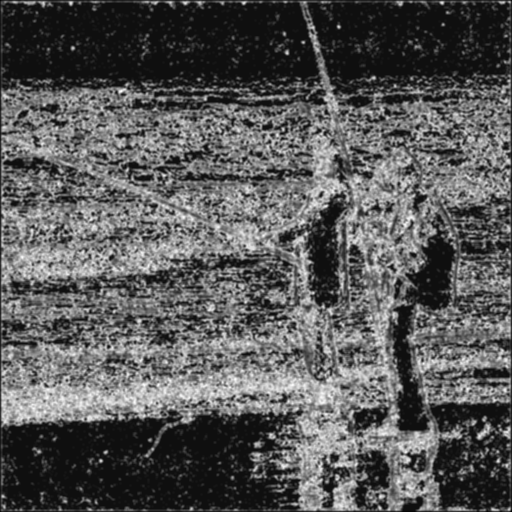}}
    \renewcommand\thesubfigure{(a\arabic{subfigure})}
    \setcounter{subfigure}{2}
    \subfigure[]{\includegraphics[width=2.5cm]{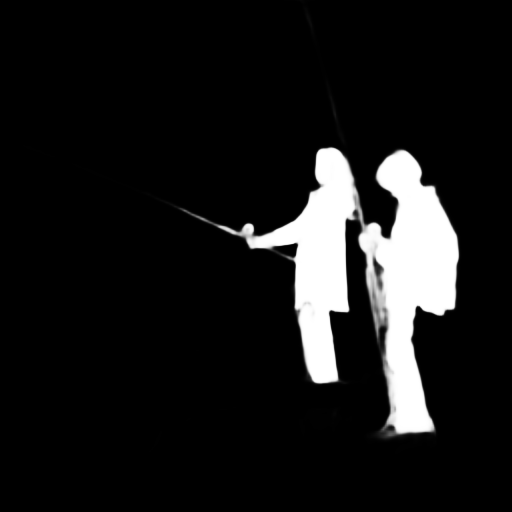}}
    \renewcommand\thesubfigure{(b\arabic{subfigure})}
    \setcounter{subfigure}{0}
    \subfigure[]{\includegraphics[width=2.5cm]{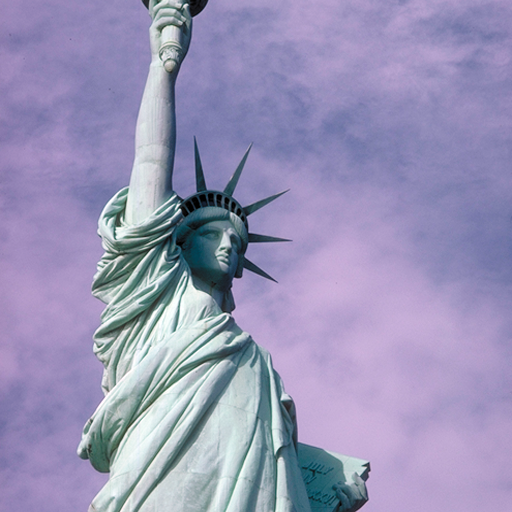}}
    \renewcommand\thesubfigure{(b\arabic{subfigure})}
    \setcounter{subfigure}{1}
    \subfigure[]{\includegraphics[width=2.5cm]{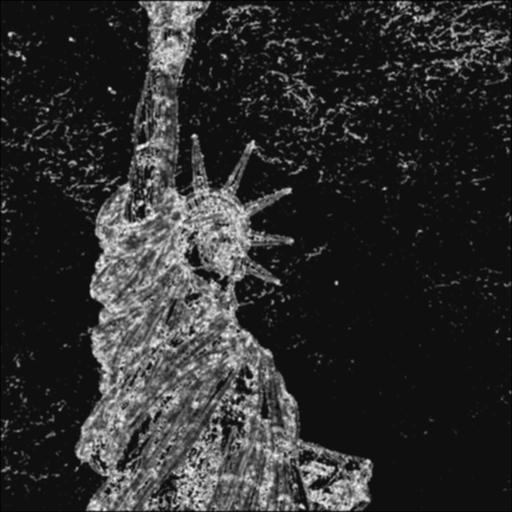}}
    \renewcommand\thesubfigure{(b\arabic{subfigure})}
    \setcounter{subfigure}{2}
    \subfigure[]{\includegraphics[width=2.5cm]{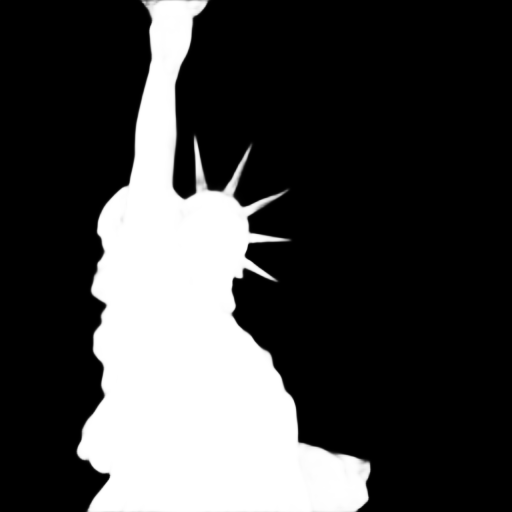}}
    
    \caption{Illustration of original images and their corresponding pattern maps as well as visual attention maps. (a1)-(b1) are original images. Their corresponding pattern complexity maps and visual attention maps are shown in (a2)-(b2) and (a3)-(b3), respectively.}
    \label{pc_va}
    \vspace{-0.4cm}
\end{figure}

In this work, the gradient $G$ is selected as a constraint of the magnitude of the proposed RGB-JND. It should be noticed that this kind of constraint is based on the full RGB channels instead of a single color channel used in \cite{wu2020unsupervised}. The constraint of the RGB-JND magnitude can be formulated as
\begin{equation}
\label{grad}
\ln \frac{ G^2 + J^2 + t_1}{2 \cdot G \cdot J + t_1},
\end{equation}
which is used to replace the first item in formula \eqref{formulation1}, we get
\begin{equation}
\label{formulation3}
\arg \min _{\vec{j}}\left(\alpha \cdot \ln \frac{ G^2 + J^2 + t_1}{2 \cdot G \cdot J + t_1}+\beta \cdot \mathcal{M}(\vec{o},\vec{d}) \right),
\end{equation}
where $t_1$ is a constant to avoid the denominator being zero. During optimization, the first item achieves the minimum when $J$ is as close as $G$. Namely, it guarantees that the magnitude of the generated RGB-JND becomes proportional to the gradient of the image. In other words, formula \eqref{formulation3} achieves the quantificationally modelling of the RGB-JND magnitude according to the visual content in RGB three channels. Besides, the pattern complexity map used in \cite{wu2020unsupervised} is also used as the input in this work as shown in Fig. \ref{pc_va} (a2) and (b2), since it also reflects the complexity of the visual content in some degree, which is helpful during RGB-JND generation.  

\subsection{Adaptive IQA combination (AIC)}
\label{AIC}
To achieve effective evaluation of the generated RGB-JND, an AIC method is developed in this subsection, which contains two main steps. Step 1: the highest-related distortion group in TABLE \ref{TableII} is regarded as the distortion group of the generative neural network caused distortion (i.e., the RGB-JND) via a well-learnt distortion group classifier. Step 2: for the confirmed distortion group in step 1, three state-of-the-art IQAs are selected and combined to evaluate the RGB-JND distorted image.  

\begin{table}
\tiny
    \centering
        \caption{\centering{\scshape Five Selected IQAs}}
    \label{tab:my_label}
    \begin{tabular}{c|c|c}
\hline 
IQA Notation &Abbreviation&Full Name\\
\hline  
$\mathcal{M}_1$ &MS-SSIM\ \cite{wang2003multiscale}&Multi-Scale Structural Similarity Index Measure\\
\hline  
$\mathcal{M}_2$ &SSIM\ \cite{wang2004image}&Structural Similarity Index Measure\\
\hline  
$\mathcal{M}_3$ &DISTS\ \cite{ding2020image}&Deep Image Structure and Texture Similarity\\
\hline  
$\mathcal{M}_4$ &GMSD\ \cite{xue2014gradient}&Gradient Magnitude Similarity Deviation\\
\hline  
$\mathcal{M}_5$ &FSIM\ \cite{zhang2011fsim}&Feature Similarity Index Measure\\
\hline 
\end{tabular}
\label{TableIII}
\vspace{-0.6cm}
\end{table}
As aforementioned, there is no IQA designed for the generative neural network caused distortion. We cannot directly use such kind of IQA to effectively evaluate the RGB-JND. However, large literature demonstrate that generative neural networks are capable of generating nearly natural visual content out of nothing, especially for the details. Hence, we assume that the RGB-JND can be any kind of noise or distortion group depending on the learning tasks. Even for a specified task, the generated noise or distortion group may be different at different training iterations, which has been demonstrated in subsection \ref{distor}. In view of this, a ResNets 
\cite{he2016deep} based classifier of image distortion group is well trained, where five typical distortion groups in TABLE \ref{TableI} are involved. The classifier is denoted by $\mathcal{C(\cdot;\cdot)}$. The index of the distortion group is denoted by $n$ ($n={1,2,3,4,5}$). The well-trained distortion group classifier always selects the most related image distortion group for the generated RGB-JND during training process. Therefore, we have
\begin{equation}
\label{classifier}
n = \mathcal{C}(\vec{d}; \psi_{CL}),
\end{equation}
where $\psi_{CL}$ is the parameter of the well-learnt classifier.
As reviewed in Section \ref{RW}-B, single IQA can only achieve good performance for limited distortion types on a certain dataset. As each distortion group in TABLE \ref{TableII} consists several relevant distortion types. For example, the distortion group $n=1$ in TABLE \ref{TableII} contains the first seven distortion types in TABLE \ref{TableI}. Therefore, several IQAs need to be combined to achieve good performance on the evaluation of a distortion group for different datasets. In view of this, we totally select five state-of-the-art IQAs in this work, which are shown in TABLE \ref{TableIII}. MS-SSIM, SSIM, DISTS, GMSD, and FSIM are denoted by $\mathcal{M}_1$, $\mathcal{M}_2$, $\mathcal{M}_3$, $\mathcal{M}_4$, and $\mathcal{M}_5$, respectively. Then, three IQAs, denoted by $Comb_n = (\mathcal{M}_x,\mathcal{M}_y,\mathcal{M}_z)$, are adaptively selected and combined from the five IQAs to make evaluations for each distortion
\begin{table}
\tiny
  \centering
       \caption{\centering{\scshape IQA Combinations of Different Distortion Groups}}

   \begin{tabular}{c|c}
\hline 
Group Index ($n$) & Combination of IQAs\\
\hline  
1 & $Comb_1$ = ($\mathcal{M}_1$, $\mathcal{M}_3$, $\mathcal{M}_4$)\\
\hline  
2 & $Comb_2$ = ($\mathcal{M}_2$, $\mathcal{M}_3$, $\mathcal{M}_4$)\\
\hline      
3 & $Comb_3$ = ($\mathcal{M}_1$, $\mathcal{M}_4$, $\mathcal{M}_5$)\\
\hline  
4 & $Comb_4$ = ($\mathcal{M}_1$, $\mathcal{M}_3$, $\mathcal{M}_4$)\\
\hline  
5 & $Comb_5$ = ($\mathcal{M}_1$, $\mathcal{M}_2$, $\mathcal{M}_4$)\\
\hline  
\end{tabular}
\label{TableIV}
\vspace{-0.6cm}
\end{table}
group. Once the distortion group $n$ is confirmed by the classifier in formula \eqref{classifier}, $Comb_n$ will be confirmed, i.e., $x,y,z$ is confirmed according to the TABLE \ref{TableIV}. Therefore, the combination of the IQAs can be regarded as the function of $n$, denoted by $\mathcal{F}(\cdot)$. Then, this process can be represented as follows
\begin{equation}
\label{mapping}
Comb_n = (\mathcal{M}_x,\mathcal{M}_y,\mathcal{M}_z) = \mathcal{F}(n).
\end{equation}

After that, three metrics are selected and combined to make the distortion evaluation. Then, the second item $\mathcal{M}$ in formula \eqref{formulation3} can be replaced with $(\mathcal{M}_x,\mathcal{M}_y,\mathcal{M}_z)$ to guarantee that the generated RGB-JND will not lead to perceptual quality decrease of the HVS. We have

\begin{equation}
\label{formulation4}
\arg \min _{\vec{j}}\left(\alpha \cdot \ln \frac{ G^2 + J^2 + t_1}{2 \cdot G \cdot J + t_1}+\beta \cdot \sum\limits_{m}^{x,y,z} \gamma_m \cdot \mathcal{M}_m(\vec{o},\vec{d}) \right),
\end{equation}  
where $m$ ($m=x,y,z$) represents the selected IQA metrics in a certain IQA combination. $\gamma_m$ is the weight of IQA metric $\mathcal{M}_m$ in the combination. It should be noticed that the scores of the selected five IQAs may be a little different even for the same distorted image. Here, we use the same pre-processing as used in \cite{ding2021comparison} to make a simple adjustment of these five IQAs.  

\begin{figure*}[htbp]
    \begin{center}
        \noindent
        \includegraphics[width=15cm]{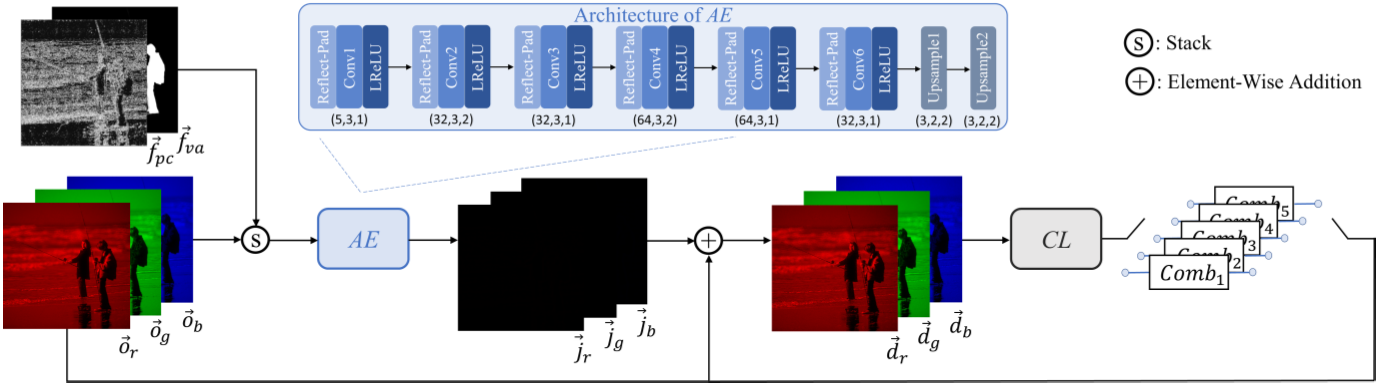}
        \caption{The framework of the proposed RGB-JND-NET. The original image $\{\vec{o}_r, \vec{o}_g, \vec{o}_b\}$ is firstly stacked with the pattern complexity map $\vec{f}_{pc}$ and visual attention map $\vec{f}_{va}$. Then, they are fed into the $AE$ module to generate the RGB-JND $\{\vec{j}_r, \vec{j}_g, \vec{j}_b\}$, where $AE$ is an auto-encoder like sub-net. Its architecture is also exhibited in this figure. $(5,3,1)$, ..., $(3,2,2)$ can be denoted by $(x,y,z)$, which represents $y \times y$ convolutions in Conv1, ..., Conv6 (or interpolations in Upsample1 and Upsample2) with $x$ filters and stride (or up sampling multiple) $z$. The reflect padding used here is set to (1, 1, 1, 1). Then, the generated RGB-JND $\{\vec{j}_r, \vec{j}_g, \vec{j}_b\}$ is added to the corresponding original image $\{\vec{o}_r, \vec{o}_g, \vec{o}_b\}$ with element-wise addition operation to generate the RGB-JND distorted image $\{\vec{d}_r, \vec{d}_g, \vec{d}_b\}$. Subsequently, the image distortion group is predicted by the well-trained $CL$ module by feeding $\{\vec{d}_r, \vec{d}_g, \vec{d}_b\}$ into $CL$, which determines the combination of IQAs $Comb_n$. Eventually, the RGB-JND distorted image is evaluated by the determined $Comb_n$ with the reference of the original image.}
    \label{FW}
    \end{center}
    \vspace{-0.6cm}
\end{figure*}

\subsection{Visual attention based spatial distribution constraint}
Large facts demonstrate that the redundancy of the HVS has relation to the visual attention. For instance, if our eyes focus on a region, the resolution of such region becomes obviously larger than that of the other regions. Besides, the resolution decreases from the focus point to the edge gradually. This causes that less noise is tolerated in the focus region compared with that in edge region. In other words, region with more visual attention has less visual redundancy, which should be assigned with smaller JND. Hence, the visual attention can be used as a good guidance on the spatial distribution of the JND. However, the visual attention is determined by lots of factors, which may be changeable. For instance, although edge region can tolerate larger noise, when the noise in such region becomes large enough, it will be noticeable by the HVS. Meanwhile, the visual attention will be reassigned and the spatial distribution of the JND will be changed as well. Therefore, the actual relationship between the visual attention and JND is more complicated than the simulated ones in the existed models as reviewed in Section \ref{RW}. It needs to be carefully modelled. However, with limitation of the knowledge on the HVS, especially for the visual attention mechanism, it's a challenge problem to build an exact relationship between the JND and the visual attention to guide and constrain the JND generation of each pixel via the HVS-inspired method.   

Besides, many works \cite{wu2020unsupervised,jin2021just} demonstrate that the relevant handcrafted features promote the performance of learning tasks, especially for the generative neural networks. Handcrafted features play an critical role in the initialization of the generative neural networks, such as the pattern complexity map and CAM map used in \cite{wu2020unsupervised,jin2021just}. In view of this, the visual attention is regarded as a handcrafted feature in this work, which will be fed into the generative neural network together with another handcrafted feature (e.g., the pattern complexity map) and original image. To facilitate the understanding, the original image and its associated pattern complexity map as well as the visual attention map are shown in Fig. \ref{pc_va}. Here, the visual attention map is obtained with the method in \cite{liu2019simple}. More details will be introduced in Section \ref{net}. The generative neural network will automatically mine the underlying relationship between the visual attention and the RGB-JND for each pixel. The relationship will guide and constrain the spatial distribution of the RGB-JND during its generation. To solve the optimization problem in formula \eqref{formulation4}, a generative neural network RGB-JND-NET will be proposed in the next subsection.

\subsection{RGB-JND-NET}
\label{net}

The framework of the proposed RGB-JND-NET contains two sub-nets ($AE$ and $CL$), two operations ($\textcircled{s}$ and $\textcircled{+}$), and an adaptive IQAs combination (AIC) evaluation module ($Comb_1,...,Comb_5$). As shown in Fig. \ref{FW}, $AE$ is an auto-encoder like sub-net to extract the useful features from the inputs and generate the reasonable RGB-JND. $CL$ is a classifier for image distortion group, which aims to classify the distortion of the RGB-JND distorted image. $\textcircled{s}$ is a channel-wise stacking operation to stack the inputs along channel dimension. $\textcircled{+}$ is a element-wise adding operation to generate the RGB-JND distorted image by adding the RGB-JND to the original image. $Comb_1,...,Comb_5$ are used for evaluating the perceptual quality of the RGB-JND distorted image with the reference of the original image.   

Unlike the characteristics of the visual content in single color channel being considered in \cite{wu2020unsupervised}, the visual content of the other two color channels and the visual attention as well as the pattern complexity are also considered by feeding them into the $AE$, namely the inputs is a $5 \times h \times w$ tensor. Besides, the architecture of auto-encoders used in this work is also different from that used in \cite{wu2020unsupervised}, more details on the architecture of the proposed $AE$ refer to the magnification part in Fig. \ref{FW}. Moreover, the proposed $AE$ is capable of generating the JND of full RGB channels (RGB-JND) instead of obtaining the JND of a single color channel in \cite{wu2020unsupervised}.

As aforementioned in subsection \ref{AIC}, $CL$ is a ResNets based classifier of image distortion group, where $CL$ is trained to well classify the five typical distortion groups as listed in TABLE \ref{TableI}. In this work, the architecture of ResNet34\footnote{https://pytorch.org/vision/0.8/\_modules/torchvision/models/resnet.html} is utilized. Also, the well-trained weights\footnote{https://download.pytorch.org/models/resnet34-333f7ec4.pth} of the ResNet34 on ImageNet \cite{krizhevsky2012imagenet} is loaded as the initialization of the $CL$. After that, $CL$ is fine-tuned on dataset TID2008 \cite{ponomarenko2009tid2008}. Once the $CL$ is well-trained, the parameters of the $CL$ will be fixed. For each iteration of the RGB-JND-NET training, the RGB-JND distorted image will be generated and fed into the $CL$ to make the distortion group classification.   

$Comb_1,...,Comb_5$ are the combinations of the IQAs, which are used to constrain that the generated RGB-JND will not lead to the decrease of the perceptual quality. Namely, $Comb_1,...,Comb_5$ are used to evaluate and supervise the RGB-JND caused perceptual quality changes. During the RGB-JND-NET training, the selection of the combination of the IQAs is determined by the result of $CL$ for each training iteration.   

The pipeline of the RGB-JND-NET is elaborated as follows. Firstly, the original image $\{\vec{o_r}, \vec{o_g}, \vec{o_b}\}$ is stacked with the pattern complexity map $\vec{f}_{pc}$ ($\vec{f}_{pc} \in \mathbb{R} ^{1 \times h \times w}$) and visual attention map $\vec{f}_{va}$ ($\vec{f}_{va} \in \mathbb{R} ^{1 \times h \times w}$) along the channel dimension, as shown in Fig. \ref{FW}. This process can be represented as

\begin{equation}
\label{stack}
\{\vec{o}_r, \vec{o}_g, \vec{o}_b, \vec{f}_{pc}, \vec{f}_{va}\} =  \{\vec{o}_r, \vec{o}_g, \vec{o}_b\} \ \textcircled{s}\ \{\vec{f}_{pc}, \vec{f}_{va}\}.
\end{equation}  

Then, the stacked features $\{\vec{o}_r, \vec{o}_g, \vec{o}_b, \vec{f}_{pc}, \vec{f}_{va}\}$ become a $5 \times h \times w$ tensor and are fed into the $AE$, denoted by $\mathcal{A}(\cdot;\cdot)$. After that, the RGB-JND $\{\vec{j}_r, \vec{j}_g, \vec{j}_b\}$ is obtained

\begin{equation}
\label{jnd generation}
\{\vec{j}_r, \vec{j}_g, \vec{j}_b\} =  \mathcal{A} \left( \{\vec{o}_r, \vec{o}_g, \vec{o}_b, \vec{f}_{pc}, \vec{f}_{va}\}; \psi_{AE} \right),
\end{equation} 
where $\psi_{AE}$ is the parameter of the sub-net $AE$. Then, the element-wise addition is applied on the RGB-JND and original image. Afterwards, the RGB-JND distorted image $\{\vec{d}_r, \vec{d}_g, \vec{d}_b\}$ is obtained as formulated in formula \eqref{distorted}. It should be noticed that such operation is performed along the channel dimension.

Then, the RGB-JND distorted image is fed into the $CL$ and classified into a certain image distortion group $n$, as formulated in formula \eqref{classifier}. Subsequently, the combination of the IQAs is determined by $n$ as shown in TABLE \ref{TableIV}. Finally, the RGB-JND distorted image is measured by the IQAs combination with the reference of its associated original image. 

In the proposed RGB-JND-NET, the loss function $\mathcal{L}$ used for optimizing the RGB-JND-NET contains two sub-losses, as formulated in \eqref{formulation4} in subsection \ref{model}-D. Therefore, we have

\begin{equation}
\label{loss}
\left\{
\begin{aligned}
&\mathcal{L} = \alpha \cdot \mathcal{L}_1 + \beta \cdot \mathcal{L}_2, \\
&\mathcal{L}_1 = \ln \frac{ G^2 + J^2 + t_1}{2 \cdot G \cdot J + t_1},\\
&\mathcal{L}_2 = \sum\limits_{m}^{x,y,z} \gamma_m \cdot \mathcal{M}_m (\{\vec{o}_r, \vec{o}_g, \vec{o}_b\},\{\vec{d}_r, \vec{d}_g, \vec{d}_b\}),
\end{aligned}
\right.
\end{equation}
where $\mathcal{L}_1$ is the gradient based magnitude constraint of the RGB-JND generation. $\mathcal{L}_2$ is the perceptual quality constraint of the generated RGB-JND. It should be noticed that the $\psi_{AE}$ in the $AE$ are the parameters to be optimized during the training. For the parameters in the $CL$, they have been fixed after well fine-tuning in the TID2008 dataset. 

By using the proposed RGB-JND-NET above, the characteristics of the full RGB channels are fully used. For instance, the visual contents in RGB three channels (original image) are fed as the inputs instead of that in the single one used in \cite{wu2020unsupervised}, which provide additional information of the other two channels for the RGB-JND generation. Meanwhile, the visual content changes in RGB three channels are also fully evaluated in the AIC to guarantees that the generated RGB-JND is the result of taking the impact of stimuli in full RGB channels into account. All these designs aim to make the RGB-JND-NET highly consistent with the perceptual of HVS. With such kinds of designs, we can directly generate the accurate JND for full RGB channels.

\begin{figure*}[htbp]
    \renewcommand\thesubfigure{}
    \centering
    \subfigure{\includegraphics[width=16cm]{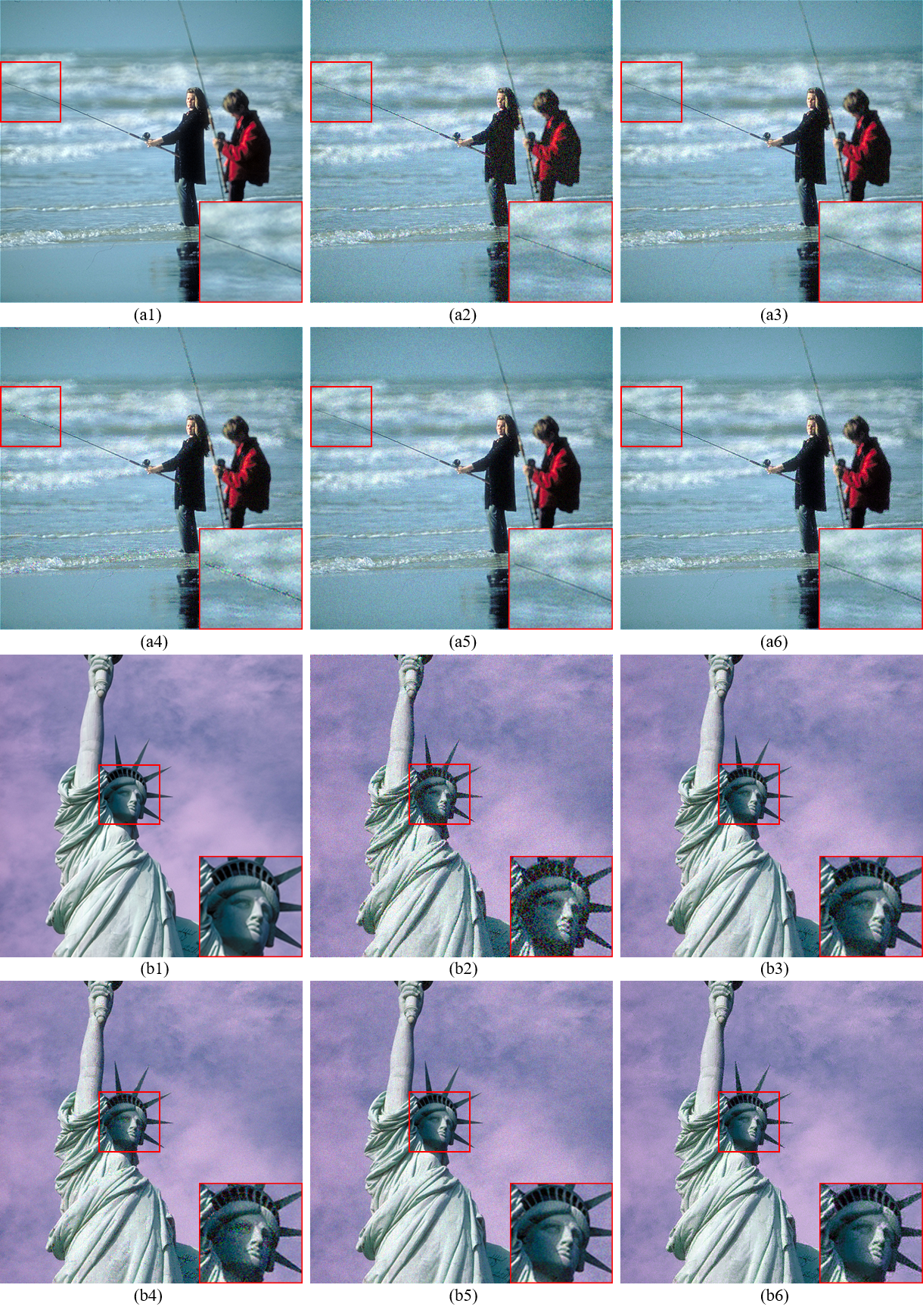}}
    \caption{Comparison of the images distorted by different JND models. (a1)-(b1) are the original images. (a2)-(b2), (a3)-(b3), (a4)-(b4), (a5)-(b5), and (a6)-(b6) are the images distorted by the JND model of Liu2010 \cite{liu2010just}, Wu2013 \cite{wu2013just}, Wu2017 \cite{wu2017enhanced}, Wu2020 \cite{wu2020unsupervised}, and ours.}
    \label{fig:my_label}
\end{figure*}


\section{Experiments}
\label{secIV}
\subsection{Datasets and evaluation metrics}
\label{secIV_A}
\subsubsection{Dataset}
Three datasets are involved in this work, i.e., TID2008 \cite{ponomarenko2009tid2008}, COCO2017 \cite{lin2014microsoft} and CSIQ \cite{larson2010categorical}. TID2008 is used to fine-tune the $CL$ before it is integrated into the RGB-JND-NET. Besides, 1000 
images are randomly selected from COCO dataset to train 
the RGB-JND-NET. Additionally, the well-trained RGB-JND-NET is tested on CSIQ dataset. It should be noticed that all the images in such dataset are cropped into $176 \times 176$ before they are fed into networks for training. 
\subsubsection{Metrics}
The five listed IQAs in TABLE \ref{TableIII} are only used for training process. Six additional IQAs are used for objective testing, which includes the Most Apparent Distortion (MAD) \cite{larson2010most}, Normalized Laplacian Pyramid Distance (NLPD) \cite{laparra2016perceptual}, Learned Perceptual Image Patch Similarity (LPIPS) \cite{zhang2018unreasonable}, Complex Wavelet SSIM (CW-SSIM) \cite{wang2005translation}, Visual Information Fidelity (VIF) \cite{sheikh2006image}, and Visual Saliency Induced (VSI) \cite{zhang2014vsi}. Besides, to facilitate network training and testing, the PyTorch implementations \cite{ding2021comparison} of eleven IQAs above are used. All the subjective test experiments are made according to the guidance of ITU-R BT.500-11 standard \cite{bt2002methodology}.

\subsection{Implementation details}
\label{secIV_B}
We compare our method with four previous state-of-the-arts JND methods, which are denoted by Liu2010 \cite{liu2010just}, Wu2013 \cite{wu2013just}, Wu2017 \cite{wu2017enhanced}, and Wu2020 \cite{wu2020unsupervised}, respectively. The first three methods are HVS-inspired models, while the last one is learning-based one. To make a fair comparison, JND images $\vec{j}$ are firstly generated via different JND models. Then, their associated JND distorted images $\vec{d}$ are generated by injecting random noise into original image with the guidance of $\vec{j}$. This process can be formulated as follows

\begin{equation}
\label{adjust}
\vec{d} = \vec{o}\ \textcircled{+}\ (\varepsilon \cdot \vec{r} \cdot \vec{j}),
\end{equation}
where $\varepsilon$ is the JND noise level adjuster. It keeps the injected noise from different models at a same level (e.g., the same PSNR or MSE). $\vec{r}$ is a random matrix with only positive 1 and negative 1 element, $\vec{r} \in \mathbb{R}^{c \times h \times w}$. In the following subsections, noise is injected in this way.

Besides, during training phase, ADAM \cite{kingma2014adam} is used to optimize the RGB-JND-NET. The learning rate is set to $10^{-5}$. The batch size is 32. $\gamma_m$, $\alpha$ and $\beta$ in formula \eqref{loss} are set to 1/3, 0.1 and 1, respectively. 
\subsection{JND models comparison}
\label{comparison}
In this subsection, we firstly compare the details of the distorted images caused by the anchor models and the proposed full RGB JND model. Then, these distorted images are evaluated by a subjective viewing test. Finally, six typical IQAs are selected to make objective tests.

\subsubsection{Details comparison}
The proposed model is firstly compared with four anchor models, i.e., Liu2010 \cite{liu2010just}, Wu2013 \cite{wu2013just}, Wu2017 \cite{wu2017enhanced}, and Wu2020 \cite{wu2020unsupervised}, with adjusting into the same level noise in different JND models according to formula \eqref{adjust}. As shown in Fig. \ref{fig:my_label}, (a1)-(b1) are the original images. (a2)-(b2), (a3)-(b3), (a4)-(b4), (a5)-(b5), and (a6)-(b6) are the images distorted images by four anchor models and the proposed model with the same level noise at PSNR = 26.06 dB, respectively. In this case, $\varepsilon$ is around 10 for our proposed model. As shown in (a2)-(b2), the overall noise are obvious with Liu2010 \cite{liu2010just}, especially for the smooth region. This case is promoted at some degree in (a3)-(b3) with Wu2013 \cite{wu2013just}, especially for the smooth region. As such region is firstly masked as the order region, more noise are injected to disorder region. The disorder and order regions are further refined and replaced with the pattern complexity estimation and masking technologies in Wu2017 \cite{wu2017enhanced}. It makes more noise injected to the high complexity regions and causes obvious distortion. Instead of directly using pattern complexity as the masking, pattern complexity is used as the feature in Wu2020 \cite{wu2020unsupervised}. By auto-mining the JND via the neural networks, it achieves improved results in (a5)-(b5).
\begin{figure}
    \centering
    \includegraphics[width=1.8cm]{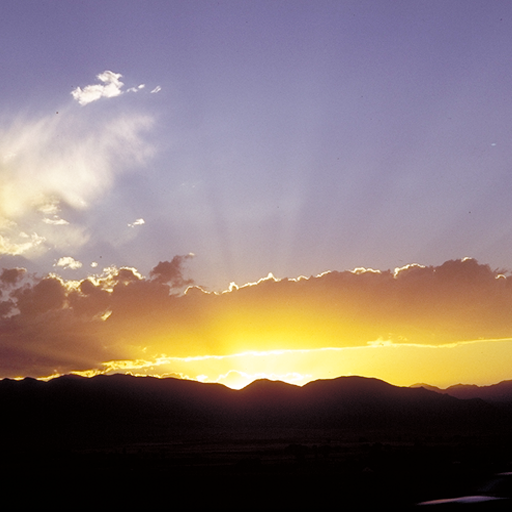}
    \hspace{0.5mm}
    \vspace{0.5mm}
    \includegraphics[width=1.8cm]{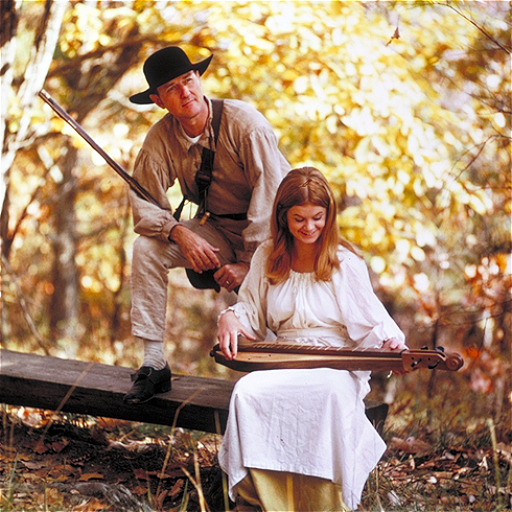}
    \hspace{0.5mm}
    \vspace{0.5mm}
    \includegraphics[width=1.8cm]{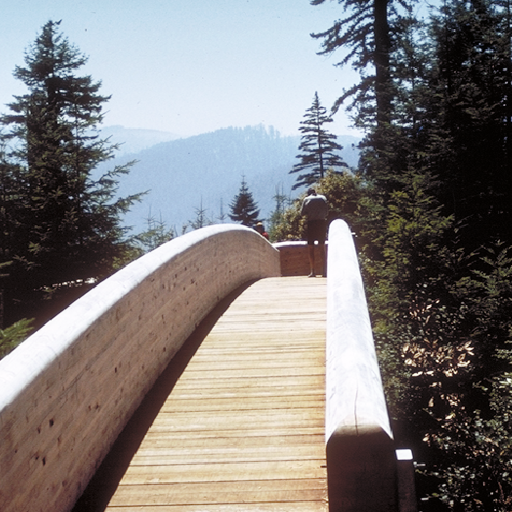}
    \hspace{0.5mm}
    \vspace{0.5mm}
    \includegraphics[width=1.8cm]{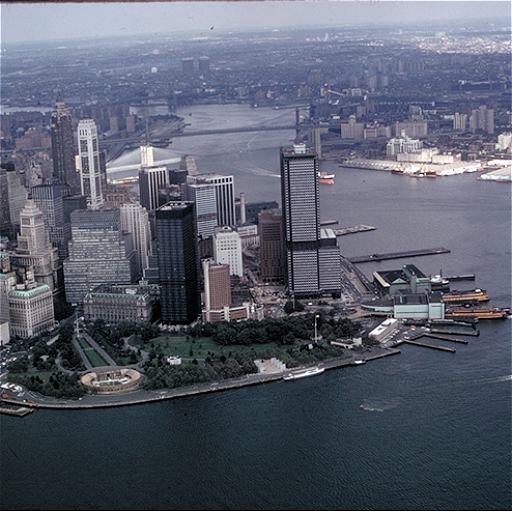}
    \hspace{0.5mm}
    \vspace{0.5mm}
    \includegraphics[width=1.8cm]{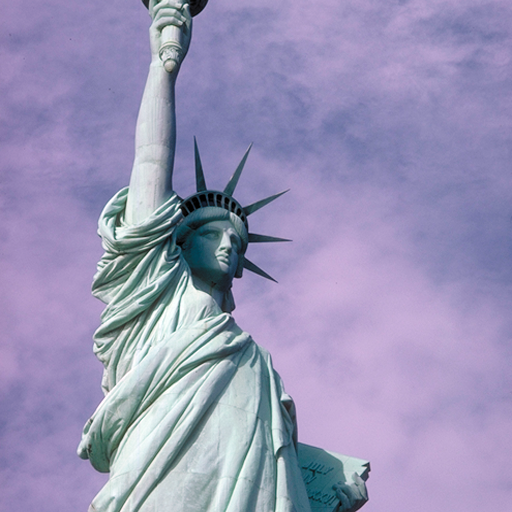}
    \hspace{0.5mm}
    \vspace{0.5mm}
    \includegraphics[width=1.8cm]{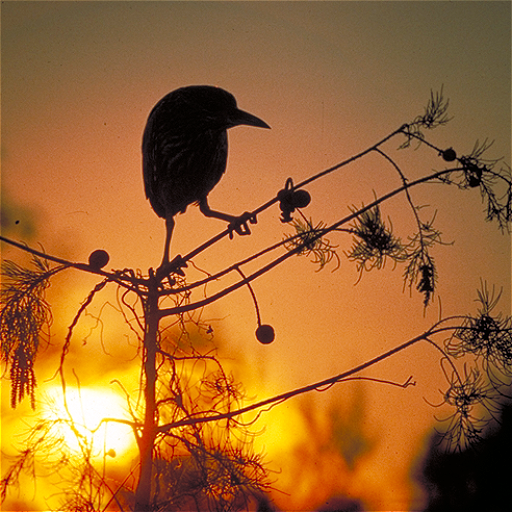}
    \hspace{0.5mm}
    \vspace{0.5mm}
    \includegraphics[width=1.8cm]{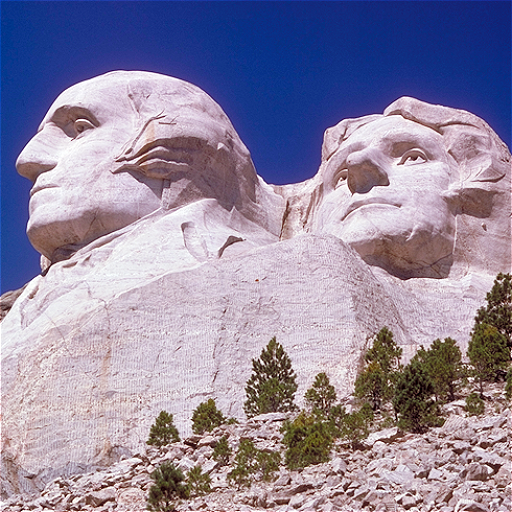}
    \hspace{0.5mm}
    \vspace{0.5mm}
    \includegraphics[width=1.8cm]{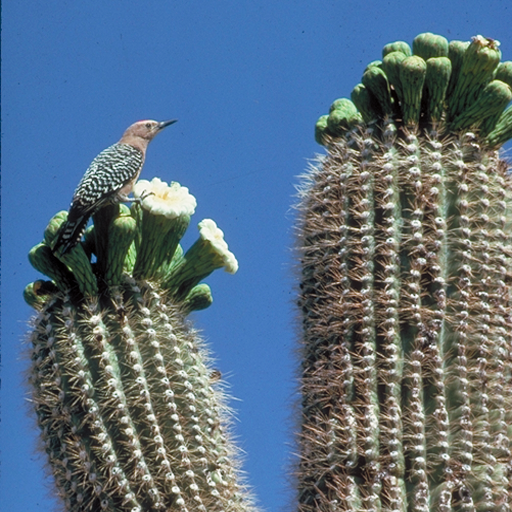}
    \hspace{0.5mm}
    \vspace{0.5mm}
    \includegraphics[width=1.8cm]{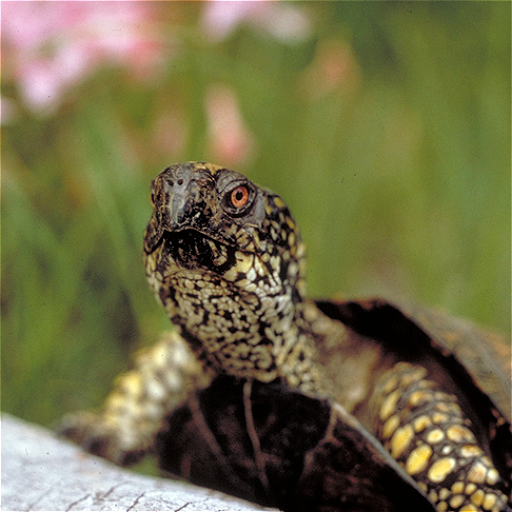}
    \hspace{0.5mm}
    \vspace{0.5mm}
    \includegraphics[width=1.8cm]{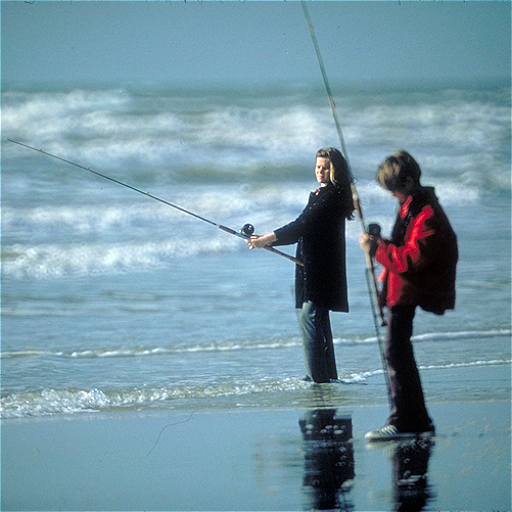}
    \hspace{0.5mm}
    \vspace{0.5mm}
    \includegraphics[width=1.8cm]{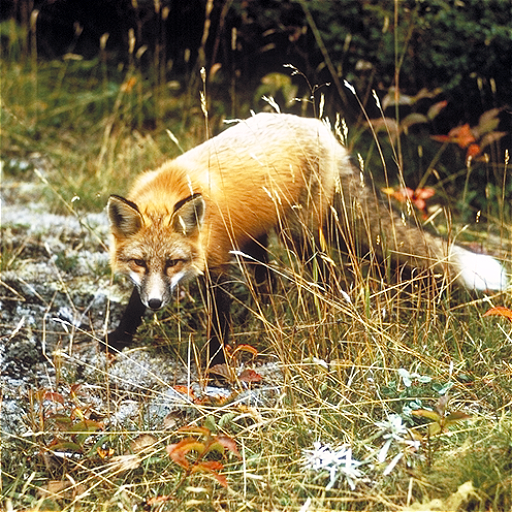}
    \hspace{0.5mm}
    \vspace{0.5mm}
    \includegraphics[width=1.8cm]{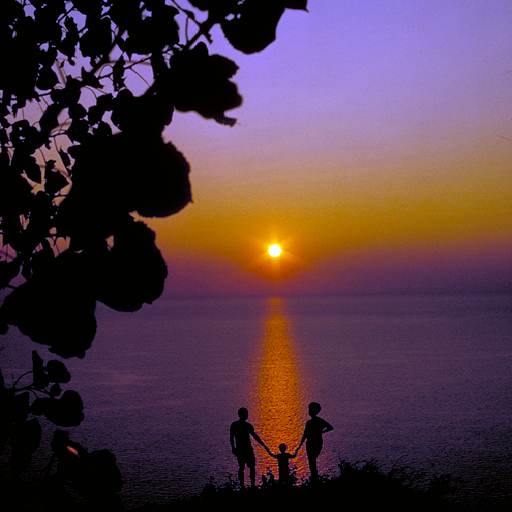}
    \caption{From left to right, from top to bottom, P1-P12 are the pictures of the objective test experiment.}
    \label{fig:CSIQ}
\vspace{-0.6cm}
\end{figure}
%
\begin{table}[htp]
\tiny
  \centering
  \caption{\centering{\scshape Experimental Evaluation Criteria And Scores of Subjective Test of Image Quality}}
  \setlength{\tabcolsep}{1.5mm}{
    \begin{tabular}{c|c|c|c|c|c|c|c|c}
    \hline
    \multicolumn{2}{c|}{\multirow{1}[2]{*}{Evaluation criterion}} & \multicolumn{3}{c|}{Left picture quality} & \multirow{1}[2]{*}{Alike} & \multicolumn{3}{c}{Right picture quality} \\
\cline{3-5}\cline{7-9}    \multicolumn{2}{c|}{} & Much better & Better & A little better &       & A little better & Better & Much better \\
    \hline
    \multicolumn{2}{c|}{Score} & 3     & 2     & 1     & 0     & -1    & -2    & -3 \\
    \hline
    \end{tabular}}%
  \label{tab:criteria}%
  \vspace{-0.6cm}
\end{table}%
\begin{table}[ht]
  \centering
  \tiny
  \caption{\centering{\scshape Subjective Evaluation of Test Results}}
    \setlength{\tabcolsep}{1.48mm}{
    \begin{tabular}{c|p{0.7cm}<{\centering}|c|p{0.7cm}<{\centering}|c|p{0.7cm}<{\centering}|c|p{0.7cm}<{\centering}|c}
    \hline
    \multicolumn{1}{c|}{\multirow{1}[2]{*}{\newline{}Index}} & \multicolumn{2}{c|}{Ours VS. Liu2010\cite{liu2010just}} & \multicolumn{2}{c|}{Ours VS. Wu2013\cite{wu2013just}} & \multicolumn{2}{c|}{Ours VS. Wu2017\cite{wu2017enhanced}} & \multicolumn{2}{c}{Ours VS. Wu2020\cite{wu2020unsupervised}} \\
\cline{2-9}    \multicolumn{1}{c|}{} &   Mean  & Std   & Mean  & Std   & Mean  & Std   & Mean  & Std \\
    \hline
    \multicolumn{1}{c|}{P1} & 0.87  & 1.03  & 1.09  & 1.21  & 1.35  & 1.17  & 1.78  & 1.32 \\
    \hline
    \multicolumn{1}{c|}{P2} & 1.35  & 1.05  & 0.65  & 0.96  & 0.87  & 1.19  & 0.65  & 1.17 \\
    \hline
    \multicolumn{1}{c|}{P3} & 1.43  & 0.88  & 0.83  & 0.92  & 0.57  & 0.88  & 1.91  & 1.02 \\
    \hline
    \multicolumn{1}{c|}{P4} & 1.57  & 1.28  & 0.65  & 0.70   & 1.00     & 1.32  & 1.52  & 1.14 \\
    \hline
    \multicolumn{1}{c|}{P5} & 1.96  & 1.60   & 1.74  & 1.11  & 1.70   & 1.27  & 1.61  & 1.01 \\
    \hline
    \multicolumn{1}{c|}{P6} & 0.91  & 0.72  & 0.26  & 0.90   & -0.13  & 0.85  & 1.48  & 0.88 \\
    \hline
    \multicolumn{1}{c|}{P7} & 1.39  & 1.52  & 0.74  & 0.99  & 0.65  & 0.81  & 0.96  & 1.20 \\
    \hline
    \multicolumn{1}{c|}{P8} & 1.48  & 0.88  & 0.96  & 0.75  & 0.57  & 0.88  & 1.96  & 1.33 \\
    \hline
    \multicolumn{1}{c|}{P9} & 1.30   & 1.23  & 0.74  & 0.90   & 0.83  & 0.56  & 1.43  & 1.41 \\
    \hline
    \multicolumn{1}{c|}{P10} & 1.39  & 1.01  & 0.78  & 0.83  & 1.04  & 1.04  & 1.35  & 1.27 \\
    \hline
    \multicolumn{1}{c|}{P11} & 0.74  & 0.79  & 0.30   & 0.75  & 0.96  & 1.16  & 1.09  & 1.25 \\
    \hline
    \multicolumn{1}{c|}{P12} & 0.04  & 1.00     & -0.09 & 0.88  & -0.48  & 1.10   & 1.65  & 0.81 \\
    \hline
    \multicolumn{1}{c|}{Average} & \textbf{\textcolor[RGB]{255,0,0}{1.20}} & — & \textbf{\textcolor[RGB]{255,0,0}{0.72}} & — & \textbf{\textcolor[RGB]{255,0,0}{0.74}} & — & \textbf{\textcolor[RGB]{255,0,0}{1.45}} & — \\
    \hline
    \end{tabular}}%
  \label{tab:mos}%
  \vspace{-0.3cm}
\end{table}
\begin{table}[ht]
   \centering
       \caption{\centering{\scshape Performance Comparisons of JND models in Terms of Different IQAs}}
      \setlength{\tabcolsep}{1.7mm}{
\tiny
\begin{tabular}{l|c|c|c|c|c|c}
\hline 
\diagbox{MODEL}{IQA} & MAD \cite{larson2010most} & NLPD \cite{laparra2016perceptual} & LPIPS \cite{zhang2018unreasonable} & CW-SSIM \cite{wang2005translation} & VIF \cite{sheikh2006image} & VSI \cite{zhang2014vsi}\\
\hline 
Liu2010 \cite{liu2010just} & 76.1780 & 0.2705 & 0.3320 & 0.9828 & 0.3886 & 0.9789\\
\hline 
Wu2013 \cite{wu2013just} & 51.7792 & 0.1941 & 0.2443 & 0.9906 & 0.4932 & 0.9898\\
\hline 
Wu2017 \cite{wu2017enhanced} & 55.6233 & 0.1908 & 0.2373 & 0.9921 & 0.5125 & 0.9901\\
\hline 
Wu2020 \cite{wu2020unsupervised} & 73.1976 & 0.2058 & 0.2728 & 0.9871 & 0.4868 & 0.9893\\
\hline 
Ours & \textbf{\textcolor[RGB]{255,0,0}{30.8754}} & \textbf{\textcolor[RGB]{255,0,0}{0.1426}} & \textbf{\textcolor[RGB]{255,0,0}{0.2297}} & \textbf{\textcolor[RGB]{255,0,0}{0.9952}} & \textbf{\textcolor[RGB]{255,0,0}{0.5918}} & \textbf{\textcolor[RGB]{255,0,0}{0.9958}}\\
\hline 
Reference & \textbf{0} & \textbf{0} & \textbf{0} & \textbf{1} & \textbf{1} & \textbf{1}\\
\hline
\end{tabular}}
\label{tab:IQA results}
\vspace{-0.6cm}
\end{table}
However, only a single IQA is used to supervise the JND generation, which is limited for various distortion evaluation during JND generation. Besides, all these methods can only achieve JND modelling on a single channel. Therefore, they estimate the JND of the RGB three color channels separately and regardless the interaction of changes among three color channels. By taking all these factors into account, an obvious improvement is achieved with our proposed model as shown in (a6)-(b6). More detailed reasons will be introduced in subsection \ref{expD}.    

\subsubsection{Subjective viewing test}
We randomly selected 12 images from the CSIQ dataset for subjective viewing test, as shown in Fig. \ref{fig:CSIQ}. The images, distorted by our proposed model and one of the anchors, are randomly exhibited on the left and right sides of the screen. The subjects are innocent about the corresponding relationship between the images and the JND models behind. The evaluation criteria and scores are shown in TABLE \ref{tab:criteria}. The subject was asked to rate the score on the two images that appeared on the screen. Seven different scores represent different results of the image quality comparison between left and right images. If the image on the left side is better than the one on the right side, the score is positive. Otherwise, the score is nonpositive. More details on the setting of the experimental environment refer to the ITU-R BT.500-11 criterion \cite{bt2002methodology}. In this experiment, 23 subjects were invited. The results of subjective viewing test are shown in TABLE \ref{tab:mos}. Here, two metrics are involved. One is the average of the scores, denoted by 'Mean'. The other is its associated standard deviation, denoted by 'Std'. Among all the average scores in the table, the positive value means that the proposed model is better than the others, the larger positive value the better. The standard deviation reflects whether the rating situation for all subjects is unified. A large standard deviation indicates that there are large differences among different subjects during viewing test. Otherwise, a small standard deviation is obtained. Here, we compare the proposed model with the anchors. As shown in TABLE \ref{tab:mos}, the average of the Mean values are positive. Besides, most of Mean values are larger than 1 and some of them even approach to 2. These demonstrate that our proposed model has fully outperformed others. 

\subsubsection{Objective IQAs test}
All the reference images are from CSIQ dataset. Their associated images distorted by different JND models are measured by six full-reference IQAs (MAD \cite{larson2010most}, NLPD \cite{laparra2016perceptual}, LPIPS \cite{zhang2018unreasonable}, CW-SSIM \cite{wang2005translation}, VIF \cite{sheikh2006image}, VSI \cite{zhang2014vsi}). The average results are listed in TABLE \ref{tab:IQA results}. For comparison, the full score of each IQA are exhibited in the last row of this table. For the first three IQAs, the closer the score is to zero, the better the JND model. While, for the rest of IQAs, the closer the score is to one, the better the JND model. The best results are highlighted with red color in this table, which concentrates on the results of the proposed model. In view of this, we can demonstrate that proposed model fully outperforms the other models during the objective IQAs test. 

We have made a wide comparison between the proposed JND model and four anchors in this subsection. All the test results demonstrate that the proposed full RGB JND model achieves the state-of-the-art results in JND modelling. In the following subsections, we will give further explanation on the advantages of the proposed model through several elaborate experiments.

\begin{figure*}
    \renewcommand\thesubfigure{(a)}
    \centering
    \subfigure[]{\includegraphics[width=8cm]{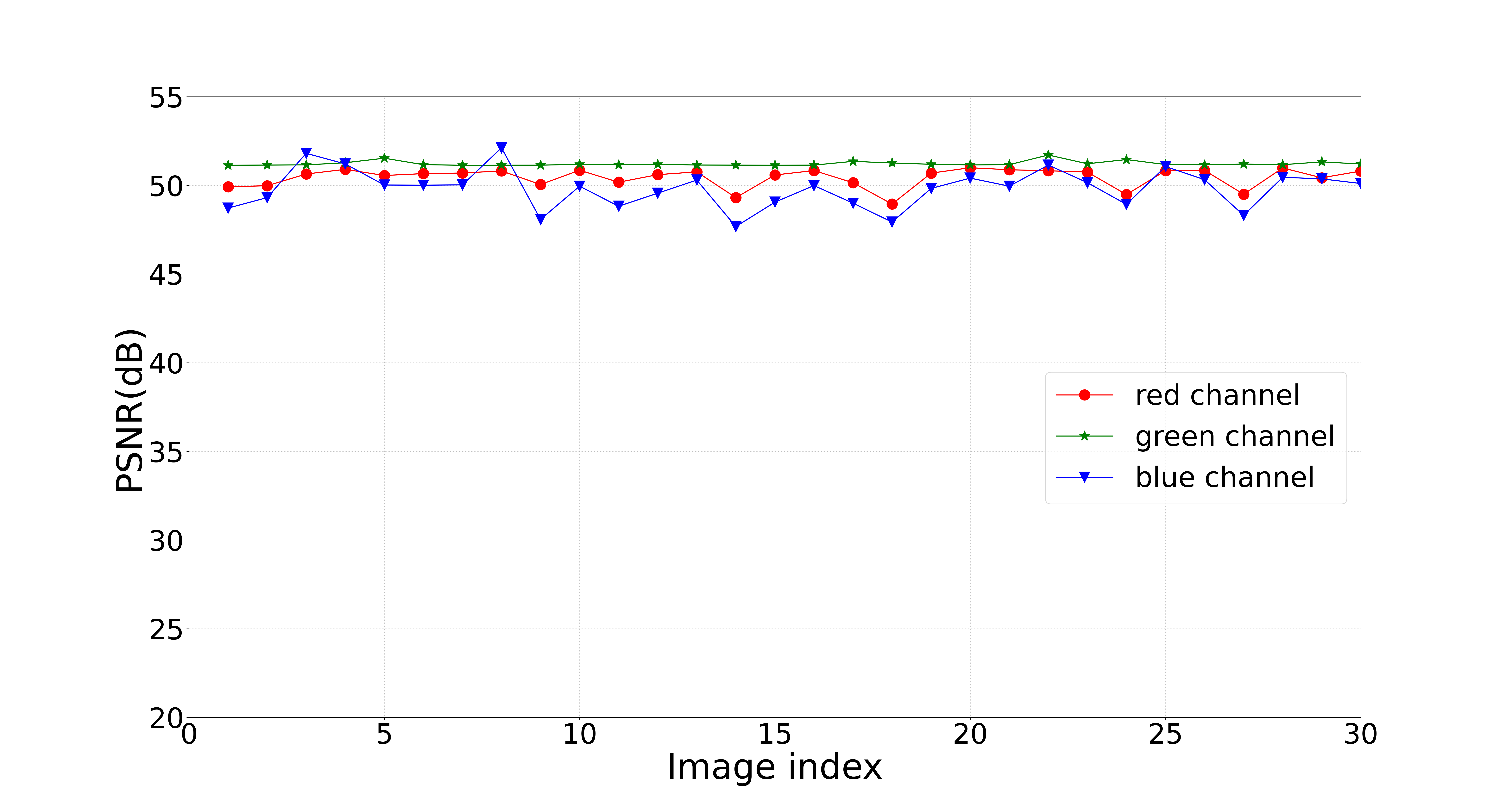}}
    \quad
    \renewcommand\thesubfigure{(b)}
    \subfigure[]{\includegraphics[width=8cm]{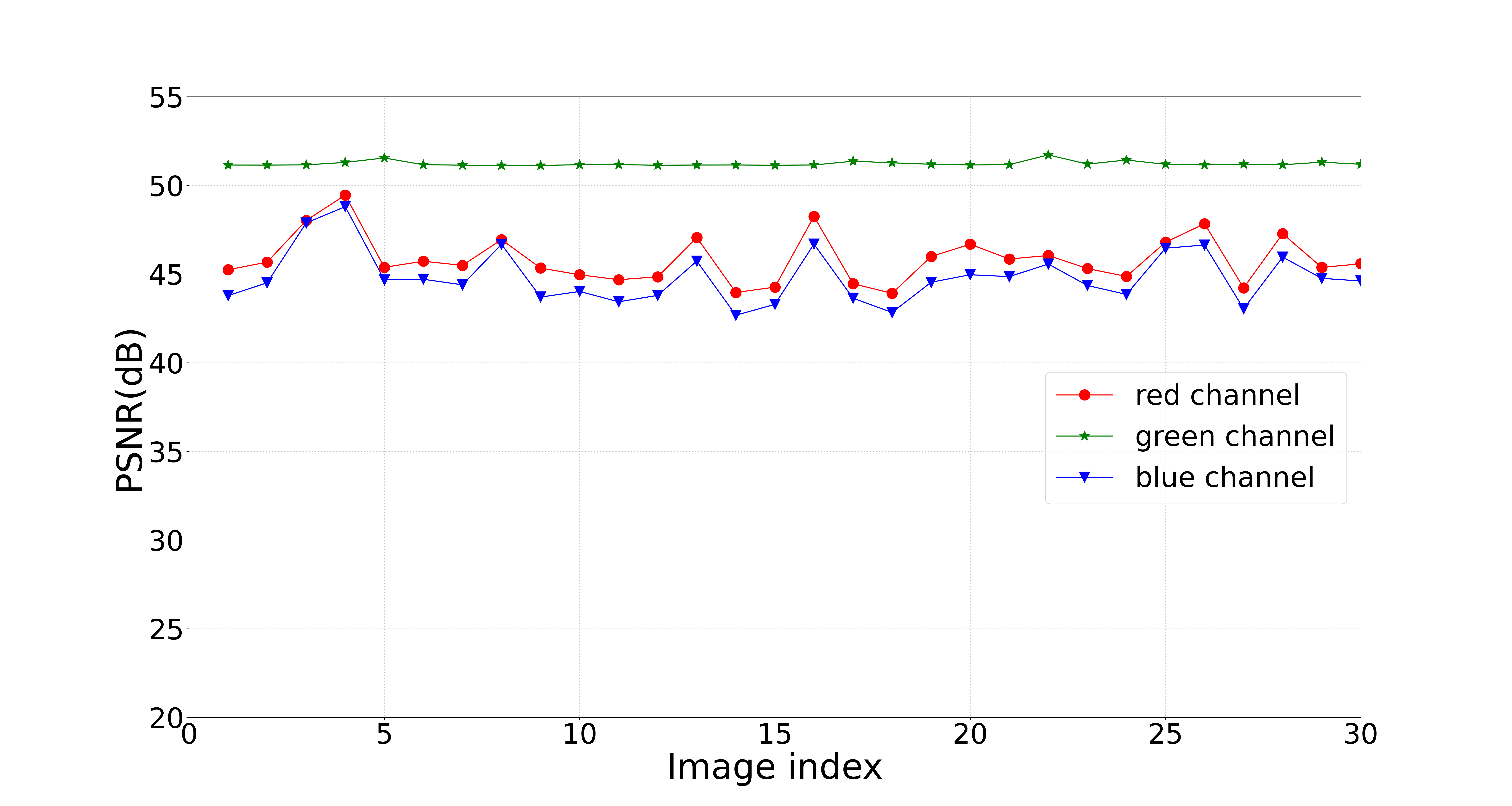}}
    \renewcommand\thesubfigure{(c)}
    \centering
    \subfigure[]{\includegraphics[width=8cm]{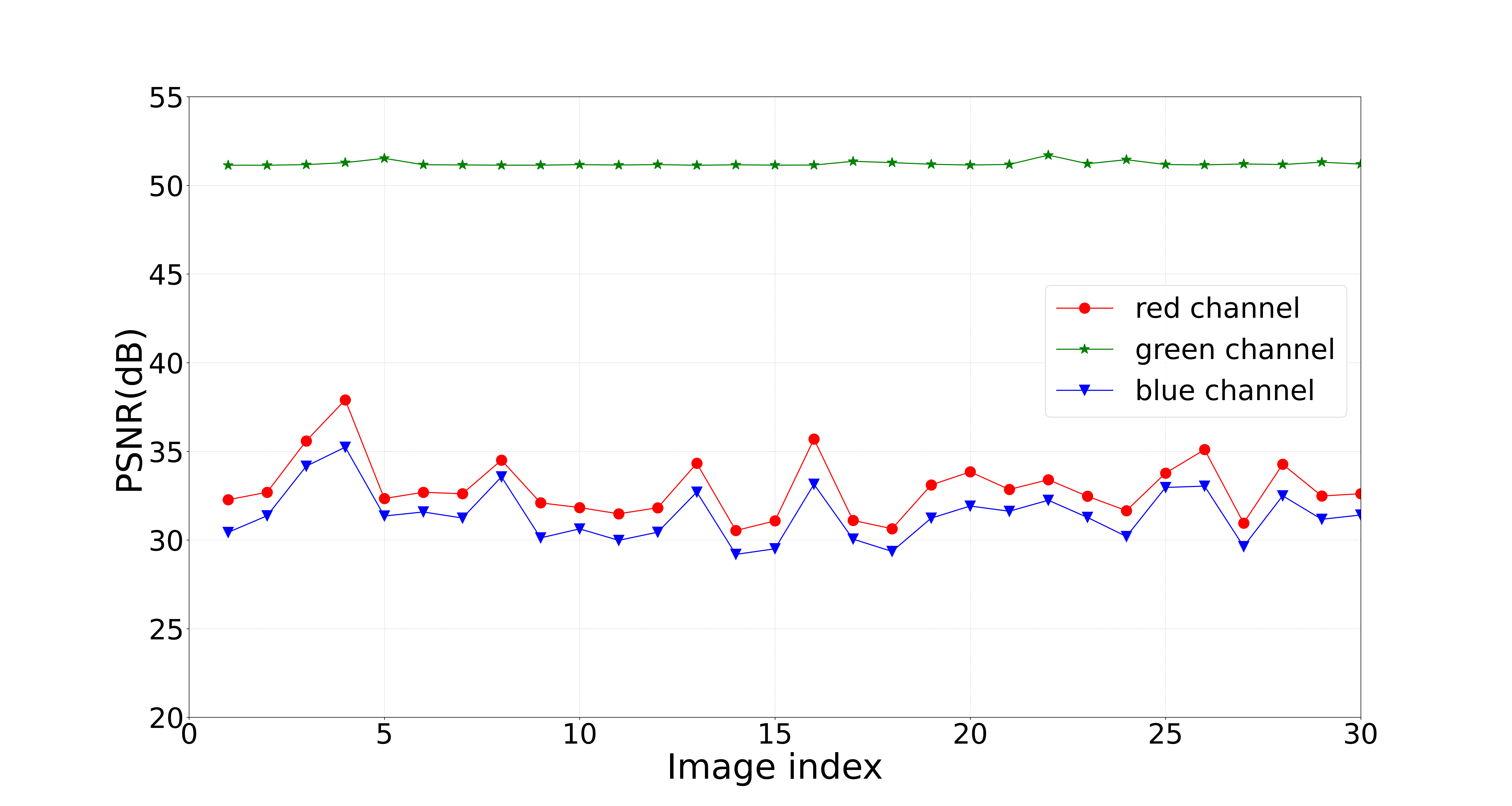}}
    \quad
    \renewcommand\thesubfigure{(d)}
    \subfigure[]{\includegraphics[width=8cm]{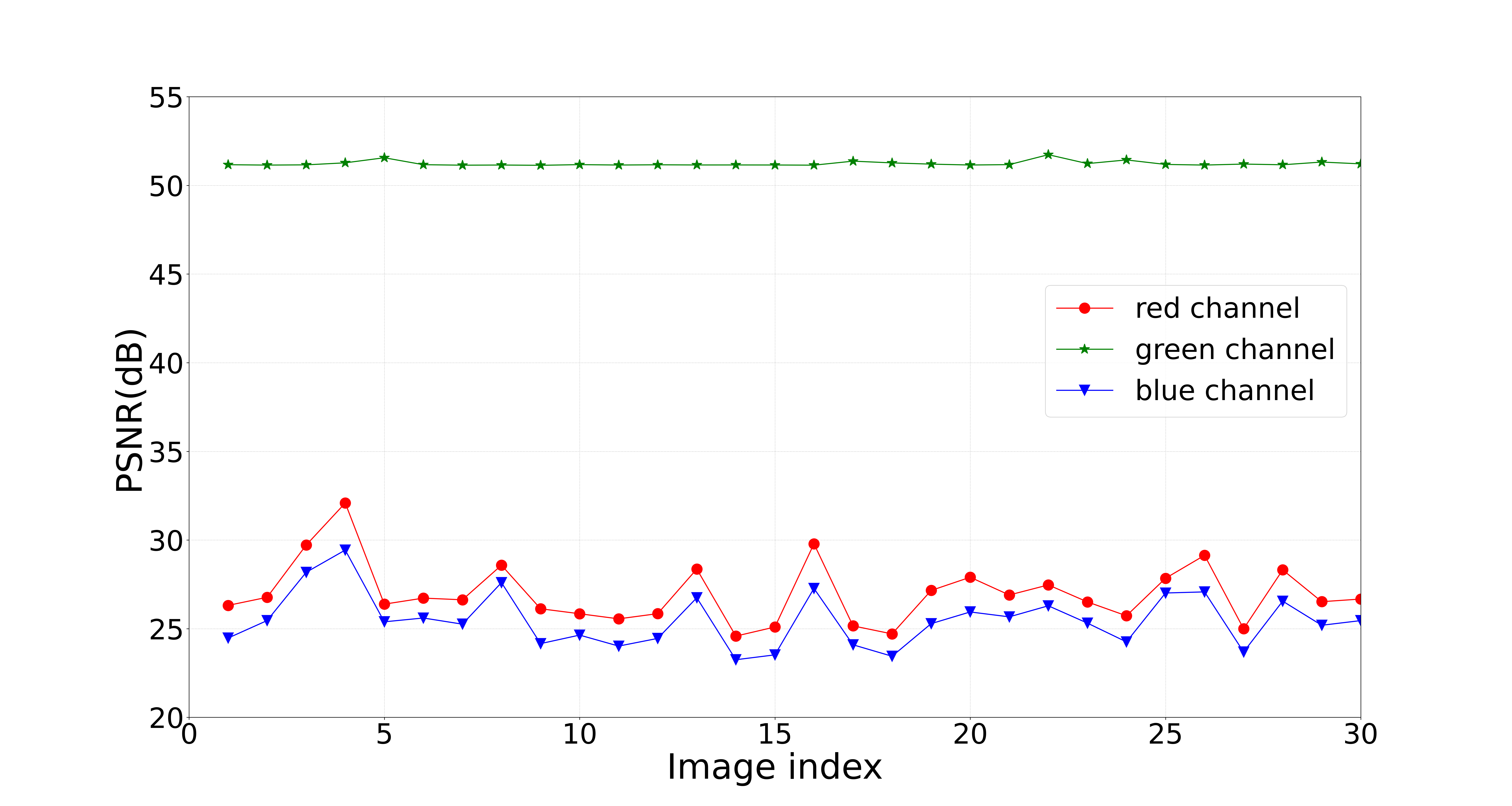}}
    \caption{PSNR values for red, green and blue channels. (a), (b), (c) and (d) are the PSNR results, when $\varepsilon$ is set to 0.5, 1, 5 and 10, respectively.}
    \label{fig:psnr}
    \vspace{-0.5cm}
\end{figure*}

\begin{figure}[h]
    \centering
    \renewcommand\thesubfigure{ }
    \renewcommand\thesubfigure{(a\arabic{subfigure})}
    \setcounter{subfigure}{0}
    \subfigure[]{\includegraphics[width=2.5cm]{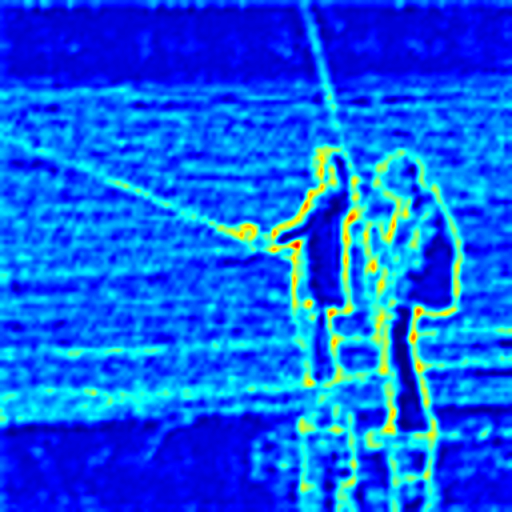}}
        \renewcommand\thesubfigure{(a\arabic{subfigure})}
    \setcounter{subfigure}{1}
    \subfigure[]{\includegraphics[width=2.5cm]{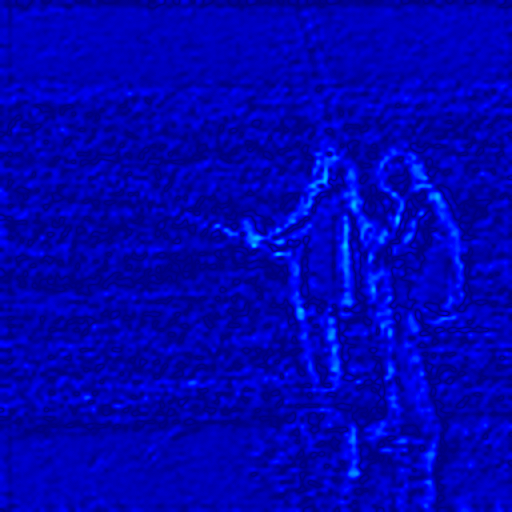}}
        \renewcommand\thesubfigure{(a\arabic{subfigure})}
    \setcounter{subfigure}{2}
    \subfigure[]{\includegraphics[width=2.5cm]{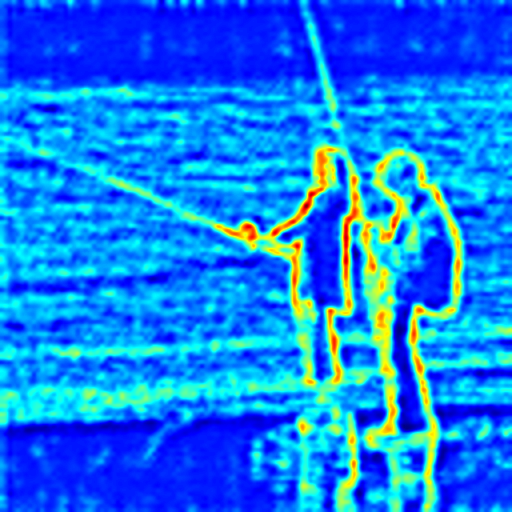}}
    \renewcommand\thesubfigure{(b\arabic{subfigure})}
    \setcounter{subfigure}{0}
    \subfigure[]{\includegraphics[width=2.5cm]{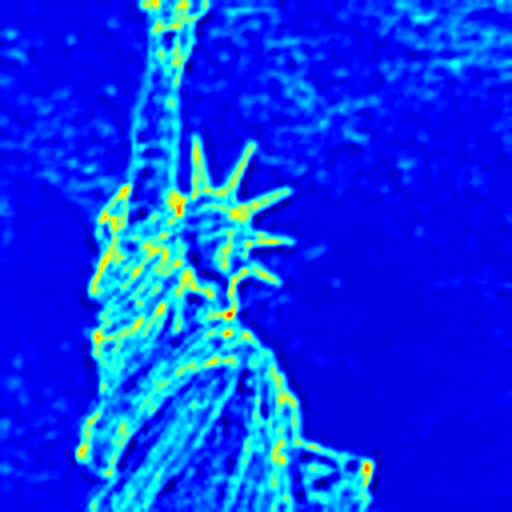}}
    \renewcommand\thesubfigure{(b\arabic{subfigure})}
    \setcounter{subfigure}{1}
    \subfigure[]{\includegraphics[width=2.5cm]{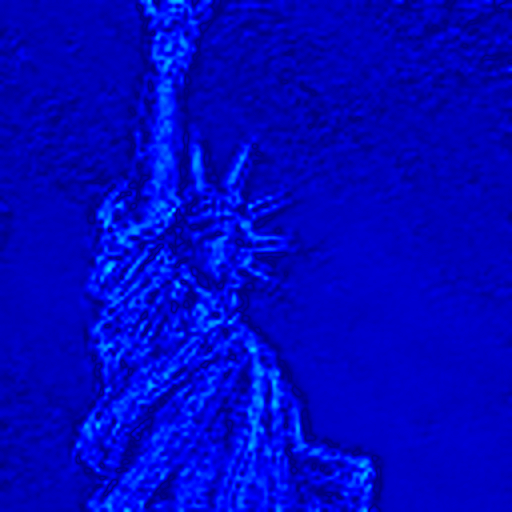}}
    \renewcommand\thesubfigure{(b\arabic{subfigure})}
    \setcounter{subfigure}{2}
    \subfigure[]{\includegraphics[width=2.5cm]{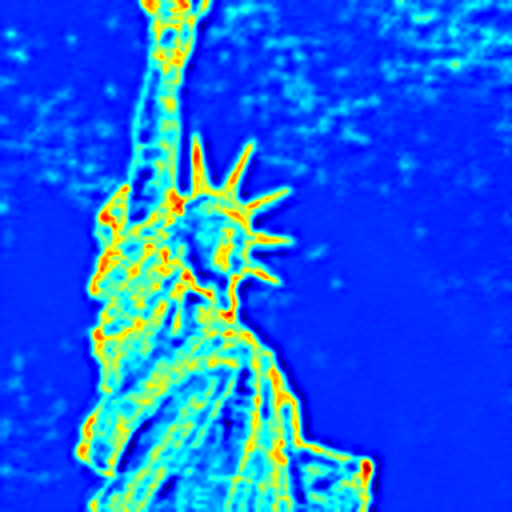}}
    \caption{Heat maps of the JND difference between each two color channels. (a1)-(b1), (a2)-(b2), and (a3)-(b3) are the heat maps of JND difference between red-green, red-blue, and green-blue channels, respectively.}
    \label{fig:diff}
    \vspace{-0.6cm}
\end{figure}

\subsection{RGB-JND distribution in three color channels}
\label{expD}

In this section, we first exhibit the RGB-JND distribution in RGB three channels. Then, the JND differences between each two color channels are shown. Finally, we come to a conclusion on our proposed model according to the results above, which explains its advantages.   

To reflect the RGB-JND distribution in three color channels, the PSNR between each color channel in the original image and its corresponding color channel in the image distorted by the proposed model with $\varepsilon=1$ is firstly calculated on the CSIQ dataset, as shown in Fig. \ref{fig:psnr} (b). The horizontal axis is the index of testing image. The vertical one is the PSNR of each color channel. We use red, green and blue lines to represent the PSNRs of different channels. In this case, we find that their PSNRs are significantly different for some images, such as the image with index = 1. For some images, such as the image with index = 4, the PSNRs of different channels are relatively close. This demonstrates that the RGB-JND distribution in three color channels is a little different for the images with different content. In other words, our proposed model can achieve accurately full RGB JND modelling according to the different visual content.

Besides, there is an interest phenomenon that the PSNR of the red channel is always smaller than that of the green one, while the PSNR of the blue one is the smallest. The largest PSNR difference achieves around 9dB. This means that our proposed model generates more JND noise in the blue and red channels compared with the green one. We also generate more RGB-JND distributions by setting $\varepsilon=0.5, 5, 10$, respectively. The results are exhibited in Fig. \ref{fig:psnr} (a), (c), (d). Based on the observation above, we find that the PSNR difference between different channels increases as $\varepsilon$ increases. When $\varepsilon=10$, the largest difference of PSNR achieves around 28 dB. Besides, the experimental results in subsection \ref{comparison} (where $\varepsilon$ is around 10) demonstrate that JND distribution of the proposed model can be well-tolerated by the HVS and achieves the best performance in both subjective viewing test and objective IQAs test compared with the other JND models. In other worlds, the experiment in this subsection provides us a good demonstration that the HVS has highest sensitivity of green, while has lowest sensitivity of blue. 

To further explore the difference of RGB-JND distribution along three color channels, we make a JND difference between each two color channels. In view of that the JND difference has small value, we apply a uniform scaling operation on the absolute value of difference between each two JNDs and obtain the heat maps of the JND difference as shown in Fig. \ref{fig:diff}. The blue and red color represent the smallest and highest difference, respectively. (a1) and (b1) are the heat maps of the JND difference between the red and green channels. The JND difference is more obvious between the green and blue channels as shown in (a3) and (b3). While, we get a much smaller JND difference between the red and blue channels as shown in (a2) and (b2). All these heat maps of the JND difference show a fact that the JND difference among different color channels mainly focus on the edge regions.  

\begin{figure*}
    \renewcommand\thesubfigure{}
    \centering
    \subfigure{\includegraphics[width=18.1cm]{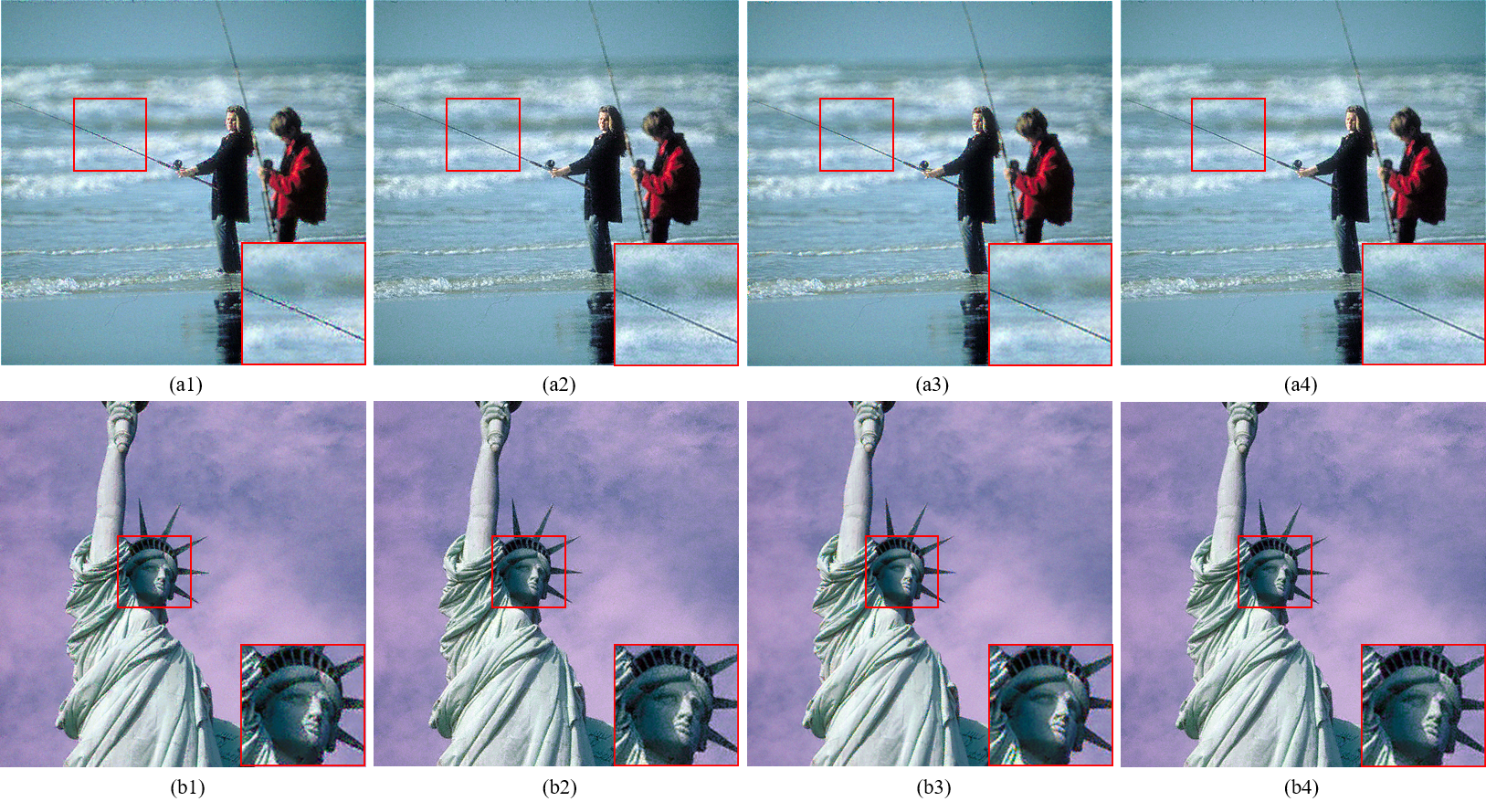}}
    \caption{Results of ablation experiment. (a1)-(b1), (a2)-(b2), (a3)-(b3), and (a4)-(b4) are the results of BL, BL + AIC, BL + VA, and BL + AIC + VA, respectively.}
    \label{fig:ablation}
    \vspace{-0.6cm}
\end{figure*}

\subsection{Ablation experiments}
In this subsection, we conduct a series of ablation experiments to identify the contribution of key components of the proposed model. In the ablation experiments, all models are trained on COCO2017 and tested on the full set of CSIQ. Except for the specified statement, all experimental setups are the same as above.

\begin{table}
  \centering
      \caption{\centering{\scshape Performance Comparisons of Ablation Experiments with Different IQAs}}
      \setlength{\tabcolsep}{1.6mm}{
\tiny
\begin{tabular}{l|c|c|c|c|c|c}
\hline 
\diagbox{MODEL}{IQA} & MAD \cite{larson2010most} & NLPD \cite{laparra2016perceptual} & LPIPS \cite{zhang2018unreasonable} & CW-SSIM \cite{wang2005translation} & VIF \cite{sheikh2006image} & VSI \cite{zhang2014vsi}\\
\hline 
BL & 52.8259 & 0.1849 & 0.2257 & 0.9926 & 0.4915 & 0.9911\\
\hline 
BL+AIC & 45.0111 & 0.1848 & 0.2545 & 0.9934 & 0.4874 & 0.9937\\
\hline 
BL+VA & 31.7818 & 0.1436 & \textcolor[RGB]{255,0,0}{\textbf{0.194}} & 0.9947 & 0.5879 & 0.9939\\
\hline 
BL+VA+AIC & \textcolor[RGB]{255,0,0}{\textbf{30.8754}} & \textcolor[RGB]{255,0,0}{\textbf{0.1426}} & {0.2297} & \textcolor[RGB]{255,0,0}{\textbf{0.9952}} & \textcolor[RGB]{255,0,0}{\textbf{0.5918}} & \textcolor[RGB]{255,0,0}{\textbf{0.9958}}\\
\hline 
Reference & \textbf{0} & \textbf{0} & \textbf{0} & \textbf{1} & \textbf{1} & \textbf{1}\\
\hline
\end{tabular}}
\label{tab:ab}
\vspace{-0.6cm}
\end{table}

There are three main components of the proposed model, i.e., gradient based magnitude constraint of full RGB channels, adaptive IQA combination (AIC), and visual attention (VA) based spatial distribution constraint. We analyze the gains of each component above by adding them into the networks gradually. Firstly, the networks with the first component is set to be the baseline network, termed as BL. Then, we take the AIC and VA into the BL and generate two networks, namely BL $+$ AIC and BL $+$ VA. Finally, all these three components are put together and we obtain the full version of the proposed model, termed as the BL $+$ AIC $+$ VA. Similarly, all these distorted images are adjusted to the same PSNR = 26.06 dB. To evaluate their performances, we firstly show the details of images distorted by BL, BL $+$ AIC, BL $+$ VA, and BL $+$ AIC $+$ VA, as shown in Fig. \ref{fig:ablation} (a1)-(b1), (a2)-(b2), (a3)-(b3), and (a4)-(b4), respectively. Based on the observation above, we can find that we get the worst performance with the BL. The comparable performance are obtained with the BL $+$ AIC and BL $+$ VA. While, we obtain the best performance with the BL $+$ VA $+$ AIC. Besides, we also evaluate the images distorted by such models with six aforementioned IQAs and obtain similar results, as shown in TABLE \ref{tab:ab}. 

From the ablation experiments above, we can make some insightful conclusions. Firstly, we can generate an initial JND modelling by applying the gradient based magnitude constraint of full RGB channels. After integrating the proposed AIC, all the undesired distortion can be recognized and avoided during JND generation and an obvious quality improvement of the JND distorted image is achieved. By further taking the visual attention based spatial distribution constraint into account, we get the best performance of the JND modelling for full RGB channels, e.g., JND spatial distribution is more reasonable, especially for the visual attention regions. Subsequently, one more experiment is made in the following subsection to further evaluate the AIC performance.

\subsection{Distortion group variations during training}
\label{distor}
\begin{figure}
    \centering
    \includegraphics[width=6.5cm]{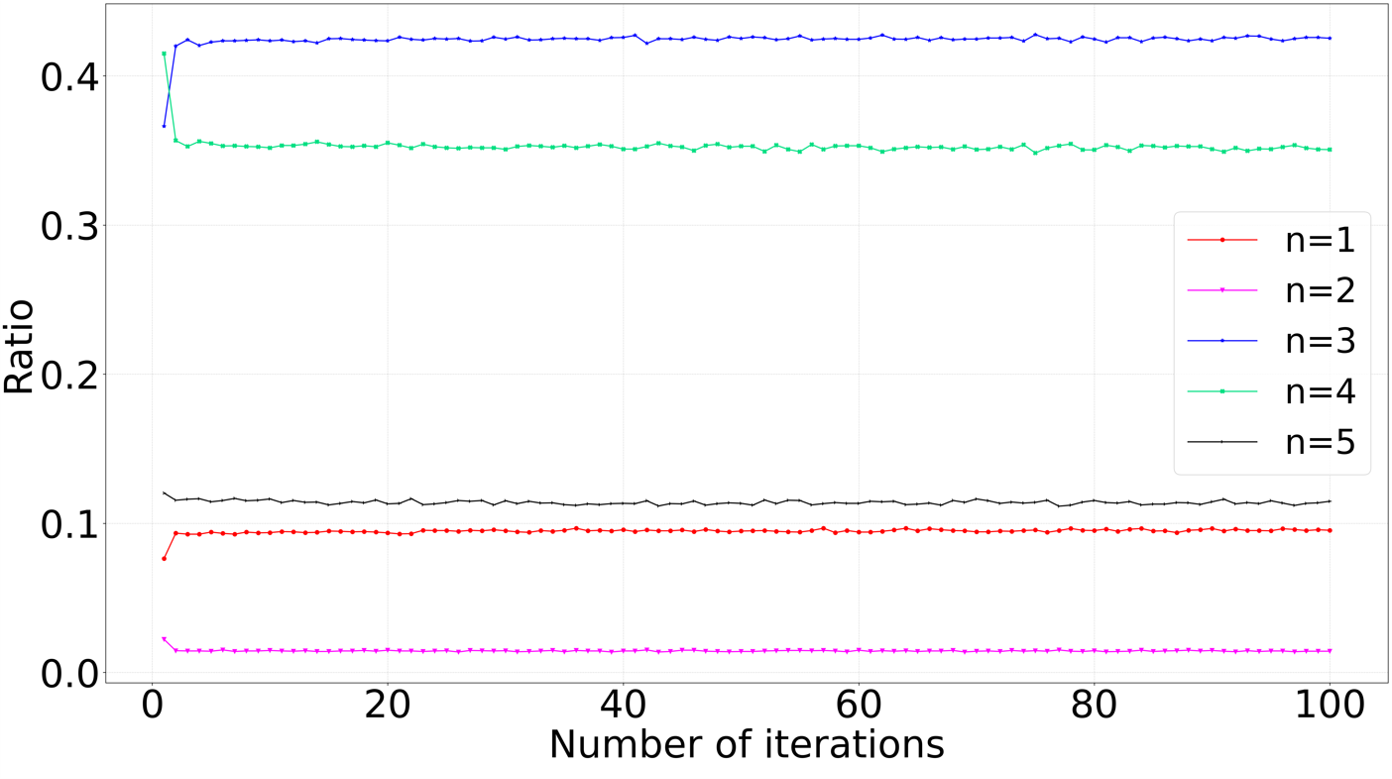}
    \caption{Variations on the ratio of different image distortion groups during training, where $n$ represents the index of image distortion group.}
    \label{fig:distortion}
\end{figure}

As aforementioned, generative neural networks may generate any kinds of distortion groups during JND generation training. Therefore, we propose the AIC to replace the single IQA (e.g., SSIM) used in \cite{wu2020unsupervised} and make an effective distortion constraint. To verify this, we collect all the image distortion groups for each iteration during the proposed model training and obtain the variation on the ratio of different image distortion groups, as shown in Fig. \ref{fig:distortion}. We use different color of lines to represent different image distortion groups. Five distortion groups ($n=1,2,...,5$) as listed in TABLE \ref{TableIV} are involved here. Based on the observation of Fig. \ref{fig:distortion}, we can find that five distortion groups change continuously with the training iteration. The ratio of each distortion group in the first few epochs changes greatly, and then becomes gently. Therefore, our proposed AIC is more reasonable, which can automatically distinguish different distortion
groups and select the best IQAs combination to constrain the JND generation. Besides, the distortion group 4 is the dominant during the first several iterations, while it is soon replaced by the distortion group 3 for the following iterations. Finally, the distortion group 3 becomes the dominant one for our generated RGB-JND. According to TABLE \ref{TableI}, the distortion group 3 refers to the distortion caused by the JPEG/JPEG2000 compression. This indicates that the generated RGB-JND is more like to the JPEG/JPEG2000 compression caused distortion in this work.

\section{Conclusion}
In this paper, we have proposed a full RGB channels JND model. Unlike mainly taking the characteristics of the luminence into account and simply regarding chrominance components as scaled versions of luminance during traditional JND modelling, this is the first work on careful investigation of the JND modelling for full-color space, where the stimuli of full-color channels are taken into account via a generative neural network, i.e., the proposed RGB-JND-NET. To make sure the generated RGB-JND can be tolerated by the HVS, an adaptive image quality assessment combination (AIC) technique is proposed. The AIC can effectively recognize and evaluate the distortion caused the by the RGB-JND-NET, which can effectively supervise the RGB-JND generation. To better constrain the RGB-JND spatial distribution, visual attention is also taken into account. Instead of being used as the rough weights of JND in most existing JND models, visual attention is regarded as a feature. RGB-JND-NET can automatically mine the underlying relationship between visual attention feature and the RGB-JND and the relationship is further used to constrain the RGB-JND spatial distribution. By applying the techniques above, the proposed model can accurately estimate the redundancy of the HVS for full RGB channels and achieves the state-of-the-art result in both subjective view test and objective evaluation. Besides, more experimental results on the RGB-JND distribution in three color channels show that the proposed model generates more JND in the red and blue channels compared with that in the green one, in line with the well-known fact that the human visual system is more sensitive to green channel in comparison with red and blue ones. Moreover, our experimental results also shows that the RGB-JND distribution in three color channels is a little different for the images with different content. In other words, our proposed model can achieve accurate full-color space JND modelling according to the visual content. All these facilitate exploring the impacts of stimuli from different color channels. JND modelling on full-color space is fundamental research, and accurate full-color JND modelling will help us to better understand the HVS. Its careful modelling is expected to catalyze positive chain-effects in color display and capture optimization, visual signal representation/compression, visual understanding, and other relevant applications.


\begin{thebibliography}{10}
	\providecommand{\url}[1]{#1}
	\csname url@samestyle\endcsname
	\providecommand{\newblock}{\relax}
	\providecommand{\bibinfo}[2]{#2}
	\providecommand{\BIBentrySTDinterwordspacing}{\spaceskip=0pt\relax}
	\providecommand{\BIBentryALTinterwordstretchfactor}{4}
	\providecommand{\BIBentryALTinterwordspacing}{\spaceskip=\fontdimen2\font plus
		\BIBentryALTinterwordstretchfactor\fontdimen3\font minus
		\fontdimen4\font\relax}
	\providecommand{\BIBforeignlanguage}[2]{{%
			\expandafter\ifx\csname l@#1\endcsname\relax
			\typeout{** WARNING: IEEEtran.bst: No hyphenation pattern has been}%
			\typeout{** loaded for the language `#1'. Using the pattern for}%
			\typeout{** the default language instead.}%
			\else
			\language=\csname l@#1\endcsname
			\fi
			#2}}
	\providecommand{\BIBdecl}{\relax}
	\BIBdecl
	
\bibitem{hall1977nonlinear}
C.~F. Hall and E.~L. Hall, ``A nonlinear model for the spatial characteristics
  of the human visual system,'' {\em IEEE Transactions on Systems, Man, and
  Cybernetics}, vol.~7, no.~3, pp.~161--170, 1977.

\bibitem{lin2021progress}
W.~Lin and G.~Ghinea, ``Progress and opportunities in modelling just-noticeable
  difference (jnd) for multimedia,'' {\em IEEE Transactions on Multimedia},
  2021.

\bibitem{wu2013perceptual}
H.~R. Wu, A.~R. Reibman, W.~Lin, F.~Pereira, and S.~S. Hemami, ``Perceptual
  visual signal compression and transmission,'' {\em Proceedings of the IEEE},
  vol.~101, no.~9, pp.~2025--2043, 2013.

\bibitem{kim2015hevc}
J.~Kim, S.-H. Bae, and M.~Kim, ``An hevc-compliant perceptual video coding
  scheme based on jnd models for variable block-sized transform kernels,'' {\em
  IEEE Transactions on Circuits and Systems for Video Technology}, vol.~25,
  no.~11, pp.~1786--1800, 2015.

\bibitem{zhou2020just}
M.~Zhou, X.~Wei, S.~Kwong, W.~Jia, and B.~Fang, ``Just noticeable
  distortion-based perceptual rate control in hevc,'' {\em IEEE Transactions on
  Image Processing}, vol.~29, pp.~7603--7614, 2020.

\bibitem{cheng2001additive}
Q.~Cheng and T.~S. Huang, ``An additive approach to transform-domain
  information hiding and optimum detection structure,'' {\em IEEE Transactions
  on Multimedia}, vol.~3, no.~3, pp.~273--284, 2001.

\bibitem{li2019orientation}
J.~Li, H.~Zhang, J.~Wang, Y.~Xiao, and W.~Wan, ``Orientation-aware saliency
  guided jnd model for robust image watermarking,'' {\em IEEE Access}, vol.~7,
  pp.~41261--41272, 2019.

\bibitem{wang2018analysis}
H.~Wang, X.~Zhang, C.~Yang, and C.-C.~J. Kuo, ``Analysis and prediction of
  jnd-based video quality model,'' in {\em 2018 Picture Coding Symposium
  (PCS)}, pp.~278--282, IEEE, 2018.

\bibitem{wang2018user}
H.~Wang, I.~Katsavounidis, X.~Zhang, C.~Yang, and C.-C.~J. Kuo, ``A user model
  for jnd-based video quality assessment: theory and applications,'' in {\em
  Applications of Digital Image Processing XLI}, vol.~10752, p.~107520M,
  International Society for Optics and Photonics, 2018.

\bibitem{cheng2018performance}
Z.~Cheng, H.~Sun, M.~Takeuchi, and J.~Katto, ``Performance comparison of
  convolutional autoencoders, generative adversarial networks and
  super-resolution for image compression,'' in {\em Proceedings of the IEEE
  Conference on Computer Vision and Pattern Recognition Workshops},
  pp.~2613--2616, 2018.

\bibitem{chou1995perceptually}
C.-H. Chou and Y.-C. Li, ``A perceptually tuned subband image coder based on
  the measure of just-noticeable-distortion profile,'' {\em IEEE Transactions
  on Circuits and Systems for Video Technology}, vol.~5, no.~6, pp.~467--476,
  1995.

\bibitem{yang2005motion}
X.~Yang, W.~Lin, Z.~Lu, E.~Ong, and S.~Yao, ``Motion-compensated residue
  preprocessing in video coding based on just-noticeable-distortion profile,''
  {\em IEEE Transactions on Circuits and Systems for Video Technology},
  vol.~15, no.~6, pp.~742--752, 2005.

\bibitem{liu2010just}
A.~Liu, W.~Lin, M.~Paul, C.~Deng, and F.~Zhang, ``Just noticeable difference
  for images with decomposition model for separating edge and textured
  regions,'' {\em IEEE Transactions on Circuits and Systems for Video
  Technology}, vol.~20, no.~11, pp.~1648--1652, 2010.

\bibitem{wu2013just}
J.~Wu, G.~Shi, W.~Lin, A.~Liu, and F.~Qi, ``Just noticeable difference
  estimation for images with free-energy principle,'' {\em IEEE Transactions on
  Multimedia}, vol.~15, no.~7, pp.~1705--1710, 2013.

\bibitem{wu2017enhanced}
J.~Wu, L.~Li, W.~Dong, G.~Shi, W.~Lin, and C.-C.~J. Kuo, ``Enhanced just
  noticeable difference model for images with pattern complexity,'' {\em IEEE
  Transactions on Image Processing}, vol.~26, no.~6, pp.~2682--2693, 2017.

\bibitem{qi2016stereoscopic}
F.~Qi, D.~Zhao, X.~Fan, and T.~Jiang, ``Stereoscopic video quality assessment
  based on visual attention and just-noticeable difference models,'' {\em
  Signal, Image and Video Processing}, vol.~10, no.~4, pp.~737--744, 2016.

\bibitem{jia2006estimating}
Y.~Jia, W.~Lin, and A.~A. Kassim, ``Estimating just-noticeable distortion for
  video,'' {\em IEEE Transactions on Circuits and Systems for Video
  Technology}, vol.~16, no.~7, pp.~820--829, 2006.

\bibitem{bae2013novel}
S.-H. Bae and M.~Kim, ``A novel dct-based jnd model for luminance adaptation
  effect in dct frequency,'' {\em IEEE Signal Processing Letters}, vol.~20,
  no.~9, pp.~893--896, 2013.

\bibitem{bae2014novel}
S.-H. Bae and M.~Kim, ``A novel generalized dct-based jnd profile based on an
  elaborate cm-jnd model for variable block-sized transforms in monochrome
  images,'' {\em IEEE Transactions on Image Processing}, vol.~23, no.~8,
  pp.~3227--3240, 2014.
  
\bibitem{niu2013visual}
Y.~Niu, M.~Kyan, L.~Ma, A.~Beghdadi, and S.~Krishnan, ``Visual saliency’s modulatory effect on just noticeable distortion profile and its application in image watermarking,'' {\em Signal Processing: Image Communication},
  vol.~28, no.~8, pp.~917--928, 2013.
  
\bibitem{hadizadeh2017saliency}
H.~Hadizadeh, A.~Rajati, and I.V. Baji{\'c}, ``Saliency-guided just noticeable distortion estimation using the normalized laplacian pyramid,'' {\em IEEE Signal Processing Letters},
  vol.~24, no.~8, pp.~1218--1222, 2017.
  

\bibitem{zeng2019visual}
Z.~Zeng, H.~Zeng, J.~Chen, J.~Zhu, Y.~Zhang, and K.-K. Ma, ``Visual attention guided pixel-wise just noticeable difference model,'' {\em IEEE Access},
  vol.~7, pp.~132111--132119, 2019.



\bibitem{liu2019deep}
H.~Liu, Y.~Zhang, H.~Zhang, C.~Fan, S.~Kwong, C.-C.~J. Kuo, and X.~Fan, ``Deep
  learning-based picture-wise just noticeable distortion prediction model for
  image compression,'' {\em IEEE Transactions on Image Processing}, vol.~29,
  pp.~641--656, 2019.

\bibitem{zhang2021deep}
Y.~Zhang, H.~Liu, Y.~Yang, X.~Fan, S.~Kwong, and C.~J. Kuo, ``Deep learning
  based just noticeable difference and perceptual quality prediction models for
  compressed video,'' {\em IEEE Transactions on Circuits and Systems for Video
  Technology}, 2021.

\bibitem{tian2021perceptual}
T.~Tian, H.~Wang, S.~Kwong, and C.-C.~J. Kuo, ``Perceptual image compression
  with block-level just noticeable difference prediction,'' {\em ACM
  Transactions on Multimedia Computing, Communications, and Applications
  (TOMM)}, vol.~16, no.~4, pp.~1--15, 2021.

\bibitem{wu2020unsupervised}
Y.~Wu, W.~Ji, and J.~Wu, ``Unsupervised deep learning for just noticeable
  difference estimation,'' in {\em 2020 IEEE International Conference on
  Multimedia \& Expo Workshops (ICMEW)}, pp.~1--6, IEEE, 2020.

\bibitem{jin2021just}
J.~Jin, X.~Zhang, X.~Fu, H.~Zhang, W.~Lin, J.~Lou, and Y.~Zhao, ``Just
  noticeable difference for deep machine vision,'' {\em IEEE Transactions on
  Circuits and Systems for Video Technology}, 2021.

\bibitem{jin2016statistical}
L.~Jin, J.~Y. Lin, S.~Hu, H.~Wang, P.~Wang, I.~Katsavounidis, A.~Aaron, and
  C.-C.~J. Kuo, ``Statistical study on perceived jpeg image quality via mcl-jci
  dataset construction and analysis,'' {\em Electronic Imaging}, vol.~2016,
  no.~13, pp.~1--9, 2016.

\bibitem{wang2016mcl}
H.~Wang, W.~Gan, S.~Hu, J.~Y. Lin, L.~Jin, L.~Song, P.~Wang, I.~Katsavounidis,
  A.~Aaron, and C.-C.~J. Kuo, ``Mcl-jcv: a jnd-based h. 264/avc video quality
  assessment dataset,'' in {\em 2016 IEEE International Conference on Image
  Processing (ICIP)}, pp.~1509--1513, IEEE, 2016.

\bibitem{wang2017videoset}
H.~Wang, I.~Katsavounidis, J.~Zhou, J.~Park, S.~Lei, X.~Zhou, M.-O. Pun,
  X.~Jin, R.~Wang, X.~Wang, {\em et~al.}, ``Videoset: A large-scale compressed
  video quality dataset based on jnd measurement,'' {\em Journal of Visual
  Communication and Image Representation}, vol.~46, pp.~292--302, 2017.

\bibitem{liu2018jnd}
X.~Liu, Z.~Chen, X.~Wang, J.~Jiang, and S.~Kowng, ``Jnd-pano: Database for just
  noticeable difference of jpeg compressed panoramic images,'' in {\em Pacific
  Rim Conference on Multimedia}, pp.~458--468, Springer, 2018.

\bibitem{liu2012image}
T.-J. Liu, W.~Lin, and C.-C.~J. Kuo, ``Image quality assessment using
  multi-method fusion,'' {\em IEEE Transactions on Image Processing}, vol.~22,
  no.~5, pp.~1793--1807, 2012.

\bibitem{goodfellow2014generative}
I.~Goodfellow, J.~Pouget-Abadie, M.~Mirza, B.~Xu, D.~Warde-Farley, S.~Ozair,
  A.~Courville, and Y.~Bengio, ``Generative adversarial nets,'' {\em Advances
  in Neural Information Processing Systems}, vol.~27, 2014.

\bibitem{gatys2016image}
L.~A. Gatys, A.~S. Ecker, and M.~Bethge, ``Image style transfer using
  convolutional neural networks,'' in {\em Proceedings of the IEEE Conference
  on Computer Vision and Pattern Recognition}, pp.~2414--2423, 2016.

\bibitem{bosse2017deep}
S.~Bosse, D.~Maniry, K.-R. M{\"u}ller, T.~Wiegand, and W.~Samek, ``Deep neural
  networks for no-reference and full-reference image quality assessment,'' {\em
  IEEE Transactions on Image Processing}, vol.~27, no.~1, pp.~206--219, 2017.

\bibitem{curcio1990human}
C.~A. Curcio, K.~R. Sloan, R.~E. Kalina, and A.~E. Hendrickson, ``Human
  photoreceptor topography,'' {\em Journal of Comparative Neurology}, vol.~292,
  no.~4, pp.~497--523, 1990.

\bibitem{wang2003multiscale}
Z.~Wang, E.~P. Simoncelli, and A.~C. Bovik, ``Multiscale structural similarity
  for image quality assessment,'' in {\em The Thrity-Seventh Asilomar
  Conference on Signals, Systems \& Computers, 2003}, vol.~2, pp.~1398--1402,
  Ieee, 2003.

\bibitem{wang2004image}
Z.~Wang, A.~C. Bovik, H.~R. Sheikh, and E.~P. Simoncelli, ``Image quality
  assessment: from error visibility to structural similarity,'' {\em IEEE
  Transactions on Image Processing}, vol.~13, no.~4, pp.~600--612, 2004.

\bibitem{sheikh2006image}
H.~R. Sheikh and A.~C. Bovik, ``Image information and visual quality,'' {\em
  IEEE Transactions on Image Processing}, vol.~15, no.~2, pp.~430--444, 2006.

\bibitem{chandler2007vsnr}
D.~M. Chandler and S.~S. Hemami, ``Vsnr: A wavelet-based visual signal-to-noise
  ratio for natural images,'' {\em IEEE Transactions on Image Processing},
  vol.~16, no.~9, pp.~2284--2298, 2007.

\bibitem{damera2000image}
N.~Damera-Venkata, T.~D. Kite, W.~S. Geisler, B.~L. Evans, and A.~C. Bovik,
  ``Image quality assessment based on a degradation model,'' {\em IEEE
  Transactions on Image Processing}, vol.~9, no.~4, pp.~636--650, 2000.

\bibitem{egiazarian2006new}
K.~Egiazarian, J.~Astola, N.~Ponomarenko, V.~Lukin, F.~Battisti, and M.~Carli,
  ``New full-reference quality metrics based on hvs,'' in {\em Proceedings of
  the Second International Workshop on Video Processing and Quality Metrics},
  vol.~4, 2006.

\bibitem{sheikh2005information}
H.~R. Sheikh, A.~C. Bovik, and G.~De~Veciana, ``An information fidelity
  criterion for image quality assessment using natural scene statistics,'' {\em
  IEEE Transactions on Image Processing}, vol.~14, no.~12, pp.~2117--2128,
  2005.

\bibitem{zhang2011fsim}
L.~Zhang, L.~Zhang, X.~Mou, and D.~Zhang, ``Fsim: A feature similarity index
  for image quality assessment,'' {\em IEEE Transactions on Image Processing},
  vol.~20, no.~8, pp.~2378--2386, 2011.

\bibitem{ponomarenko2009tid2008}
N.~Ponomarenko, V.~Lukin, A.~Zelensky, K.~Egiazarian, M.~Carli, and
  F.~Battisti, ``Tid2008-a database for evaluation of full-reference visual
  quality assessment metrics,'' {\em Advances of Modern Radioelectronics},
  vol.~10, no.~4, pp.~30--45, 2009.

\bibitem{ding2020image}
K.~Ding, K.~Ma, S.~Wang, and E.~P. Simoncelli, ``Image quality assessment:
  Unifying structure and texture similarity,'' {\em arXiv preprint
  arXiv:2004.07728}, 2020.

\bibitem{xue2014gradient}
W.~Xue, L.~Zhang, X.~Mou, and A.~C. Bovik, ``Gradient magnitude similarity
  deviation: A highly efficient perceptual image quality index,'' {\em IEEE
  Transactions on Image Processing: A Publication of the IEEE Signal Processing
  Society}, vol.~23, no.~2, pp.~684--695, 2014.

\bibitem{he2016deep}
K.~He, X.~Zhang, S.~Ren, and J.~Sun, ``Deep residual learning for image
  recognition,'' in {\em Proceedings of the IEEE Conference on Computer Vision
  and Pattern Recognition}, pp.~770--778, 2016.

\bibitem{ding2021comparison}
K.~Ding, K.~Ma, S.~Wang, and E.~P. Simoncelli, ``Comparison of full-reference
  image quality models for optimization of image processing systems,'' {\em
  International Journal of Computer Vision}, vol.~129, no.~4, pp.~1258--1281,
  2021.

\bibitem{liu2019simple}
J.-J. Liu, Q.~Hou, M.-M. Cheng, J.~Feng, and J.~Jiang, ``A simple pooling-based
  design for real-time salient object detection,'' in {\em Proceedings of the
  IEEE/CVF Conference on Computer Vision and Pattern Recognition},
  pp.~3917--3926, 2019.

\bibitem{krizhevsky2012imagenet}
A.~Krizhevsky, I.~Sutskever, and G.~E. Hinton, ``Imagenet classification with
  deep convolutional neural networks,'' {\em Advances in Neural Information
  Processing Systems}, vol.~25, pp.~1097--1105, 2012.

\bibitem{larson2010most}
E.~C. Larson and D.~M. Chandler, ``Most apparent distortion: full-reference
  image quality assessment and the role of strategy,'' {\em Journal of
  Electronic Imaging}, vol.~19, no.~1, p.~011006, 2010.

\bibitem{laparra2016perceptual}
V.~Laparra, J.~Ball{\'e}, A.~Berardino, and E.~P. Simoncelli, ``Perceptual
  image quality assessment using a normalized laplacian pyramid,'' {\em
  Electronic Imaging}, vol.~2016, no.~16, pp.~1--6, 2016.

\bibitem{zhang2018unreasonable}
R.~Zhang, P.~Isola, A.~A. Efros, E.~Shechtman, and O.~Wang, ``The unreasonable
  effectiveness of deep features as a perceptual metric,'' in {\em Proceedings
  of the IEEE Conference on Computer Vision and Pattern Recognition},
  pp.~586--595, 2018.

\bibitem{wang2005translation}
Z.~Wang and E.~P. Simoncelli, ``Translation insensitive image similarity in
  complex wavelet domain,'' in {\em 2005 IEEE International Conference on
  Acoustics, Speech, and Signal Processing, ICASSP'05}, pp.~II573--II576, 2005.

\bibitem{zhang2014vsi}
L.~Zhang, Y.~Shen, and H.~Li, ``Vsi: A visual saliency-induced index for
  perceptual image quality assessment,'' {\em IEEE Transactions on Image
  Processing}, vol.~23, no.~10, pp.~4270--4281, 2014.

\bibitem{lin2014microsoft}
T.-Y. Lin, M.~Maire, S.~Belongie, J.~Hays, P.~Perona, D.~Ramanan,
  P.~Doll{\'a}r, and C.~L. Zitnick, ``Microsoft coco: Common objects in
  context,'' in {\em European Conference on Computer Vision}, pp.~740--755,
  Springer, 2014.

\bibitem{larson2010categorical}
E.~C. Larson and D.~Chandler, ``Categorical image quality (csiq) database,''
  2010.

\bibitem{bt2002methodology}
R.~I.-R. BT, ``Methodology for the subjective assessment of the quality of
  television pictures,'' {\em International Telecommunication Union}, 2002.

\bibitem{kingma2014adam}
D.~P. Kingma and J.~Ba, ``Adam: A method for stochastic optimization,'' {\em
  arXiv preprint arXiv:1412.6980}, 2014.

\end{thebibliography}
\end{document}